# Scalable machine learning framework for multiphase identification from powder X-ray diffraction

Xinyang Tong,[1] Ethan Jin,[2] Jiahan Xu,[1] Aditya Rao,[3] Pengcen Jiang,[4]
and Nathan J. Szymanski[1,*]

## Abstract

X-ray diffraction (XRD) is the primary tool for identifying crystalline phases following synthesis, but automated phase identification remains challenging, particularly for multiphase samples with overlapping peaks and experimental artifacts. While deep-learning methods have been proposed to improve upon classical search-match algorithms, most formulate phase identification as a single closed-set classification problem, requiring one shared model to discriminate among all candidate phases. Here we introduce GALAXI, which instead decouples the identification task into independent one-versus-all binary classifiers that each specialize in recognizing a single phase. These pre-trained classifiers first narrow the search space to a small set of plausible phases, which are then evaluated through Rietveld refinement to identify the combination of phases that best explains the full diffraction pattern. On a curated set of experimental patterns, GALAXI identifies the correct phases with a micro-F1 score of 0.935, outperforming classical search-match and prior deep-learning models. The method remains robust to common experimental artifacts, including low impurity phase fractions, small crystallite size, peak shifts, sample displacement, and texture, and performs well when applied to time-resolved *in-situ* XRD data from solid-state reactions. Moreover, because the phase-specific models are independent, GALAXI can expand to large reference libraries without retraining existing models. This modular architecture enables us to train classifiers for 64,594 structures from the Crystallography Open Database and deploy them through a public web interface at https://galaxi-xrd.com.

[1] University of California, Los Angeles, Department of Materials Science & Engineering, Los Angeles, CA 90095
[2] University of California, Los Angeles, Electrical and Computer Engineering Department, Los Angeles, CA 90095
[3] University of California, Los Angeles, Mechanical and Aerospace Engineering Department, Los Angeles, CA 90095
[4] University of California, San Diego, Halıcıoğlu Data Science Institute, La Jolla, CA 92093
[*] Correspondence to nszymanski@seas.ucla.edu

## Introduction

X-ray diffraction (XRD) is a fundamental technique for the structural characterization of crystalline materials, providing the feedback required to identify phases after their synthesis.[1] Traditionally, phase identification from XRD patterns has relied heavily on peak search-match algorithms and profile fitting techniques.[2] While these classical approaches are well-established, they frequently struggle when applied to challenging experimental data. Their reliability degrades in the presence of experimental artifacts and peak overlap that are prevalent in multiphase mixtures, often requiring careful manual interpretation.[3]

To address the limitations of search-match and profile fitting, data-driven approaches leveraging machine learning (ML) have gained attention. Convolutional neural networks (CNNs), in particular, can extract invariant structural features directly from noisy diffraction data.[4] Early implementations utilized CNNs to classify patterns by crystal dimensionality and space groups,[5] and later transitioned toward the identification of specific phases. A central challenge for each task is the simulation-to-experiment gap, which has been addressed by incorporating increasingly realistic instrumental and sample-related artifacts into synthetic training data. For instance, Lee *et al.* showed that models trained entirely on synthetic powder patterns could identify phases in experimental multiphase samples.[6] Subsequent work expanded this approach through more extensive training-set augmentation, pair distribution function analysis in XRD-AutoAnalyzer,[7,8] and ensemble architectures such as Crystallography Companion Agent (XCA).[9] More recently, cross-attention architectures such as XQueryer have paired physics-guided pattern simulation with transformer-based structure identification for real-time diffractometer integration.[10] Beyond this supervised setting, the field is also moving toward generative models[11–13] which are capable of proposing entirely novel structures that match provided XRD patterns. These approaches remain comparatively nascent, however, relative to supervised classification methods that are more established for practical identification of known phases.

Despite these advances, current supervised models face a critical architectural bottleneck. They predominantly treat phase identification as a single closed-set classification problem, where the output space is restricted to a fixed set of phases defined during training. This creates a practical scaling problem, because the model must be retrained whenever new phases are added or the chemical space changes, while a truly universal classifier would need to discriminate among an enormous number of possible phases. Moreover, a shared classifier must learn features that distinguish many classes simultaneously, which can limit specialization and lead to suboptimal representations for individual classes.[14,15]

Here we introduce GALAXI (General Automated Learning and Analysis for XRD-based Identification), an ML framework that decouples multiphase identification into independent, phase-specific binary classifiers. This architecture allows the reference library to expand without retraining existing models. Each classifier is trained to detect a single phase, even at low weight fractions, with an emphasis on high recall to avoid missed detections. A mask-modulated attention mechanism helps the model focus on the most informative portions of the diffraction pattern, and the resulting candidate set is then evaluated by the automated Rietveld refinement engine DARA (Data-driven Automated Rietveld Analysis)[16] to identify the combination of phases that best explains the full pattern.

GALAXI's performance is benchmarked against four baselines: peak search-match based on the Smith/Snyder method[17] and three previously published CNNs: XRD-AutoAnalyzer,[7,8] XCA,[9] and XQueryer.[10] These are evaluated on a held-out set of experimental patterns containing both single-phase and multiphase samples. GALAXI outperforms all four, achieving 82.0% top-15 accuracy compared with 32.8–47.5% for the other methods. We additionally stress-test GALAXI against several experimental artifacts, including low weight fraction, small crystallite size, peak shift, sample displacement, and texture. As an application, GALAXI is used to identify the phases that form during solid-state synthesis reactions, characterized using high-temperature *in-situ* XRD. Because the phase-specific models can be added independently without retraining the existing library, we scale GALAXI beyond these test cases by pretraining models for 64,594 structures across the periodic table, curated from the Crystallography Open Database (COD).[18] The resulting library is deployed for community use through a public web interface at https://galaxi-xrd.com.

## Workflow

As illustrated in **Figure 1**, GALAXI consists of a model-training workflow and an inference workflow for analyzing XRD patterns. During training, candidate crystal structures are first curated to retain physically plausible phases relevant to the intended experimental conditions (Methods Section 1). These structures are then used to simulate diffraction patterns that incorporate a variety of physical and instrumental artifacts (Methods Section 2), followed by a shared preprocessing pipeline for smoothing, background removal, and weak-signal enhancement.

Each candidate phase is then assigned its own lightweight one-dimensional CNN, trained as an independent binary classifier that outputs a single probability for whether that phase is present in each pattern. Because all models use the same fixed architecture and hyperparameters, adding a new candidate phase requires training only one additional model rather than retraining the existing library. Each CNN also uses a mask-modulated spatial attention mechanism that emphasizes informative diffraction features and allows the model to operate on experimental patterns with different angular ranges. Details for preprocessing, CNN model architecture, and mask-modulated attention are described in Methods Section 3.

We systematically tuned the pattern-generation protocol and model architecture against a dedicated 57-pattern experimental validation set (Supplementary Note S3), explicitly optimizing the CNNs for high recall. False negatives are strongly penalized because any phase missed by the CNN cannot be recovered downstream, whereas false positives are less costly because Rietveld refinement performed with DARA[16] can subsequently reject candidates that do not fit the pattern well (Supplementary Note S1). Tuning on this validation set was also used to optimize the preprocessing pipeline and the mask-modulated attention mechanism. The full hyperparameter and pattern-generation sweeps are reported in Supplementary Figure S1 and Table S1, with the specific contribution of the mask-modulated attention mechanism examined in Supplementary Note S3.5.

During inference, the known chemical space is first used to define a set of plausible candidate phases, drawn from the COD in this work. Candidates are restricted to phases whose constituent elements are a subset of the specified chemical space. For example, specifying Li-Mn-O includes models for Li-Mn-O ternaries, Li-O and Mn-O binaries, and the relevant elemental Li and Mn phases. The supplied pattern is then preprocessed using the same pipeline applied during training, and the corresponding phase-specific CNNs independently screen the candidate set for phases that may be present. Positively identified candidates are passed to DARA (Methods Section 4), which evaluates combinations of these phases against the full diffraction pattern to determine the final phase assignments and quantitative weight fractions.

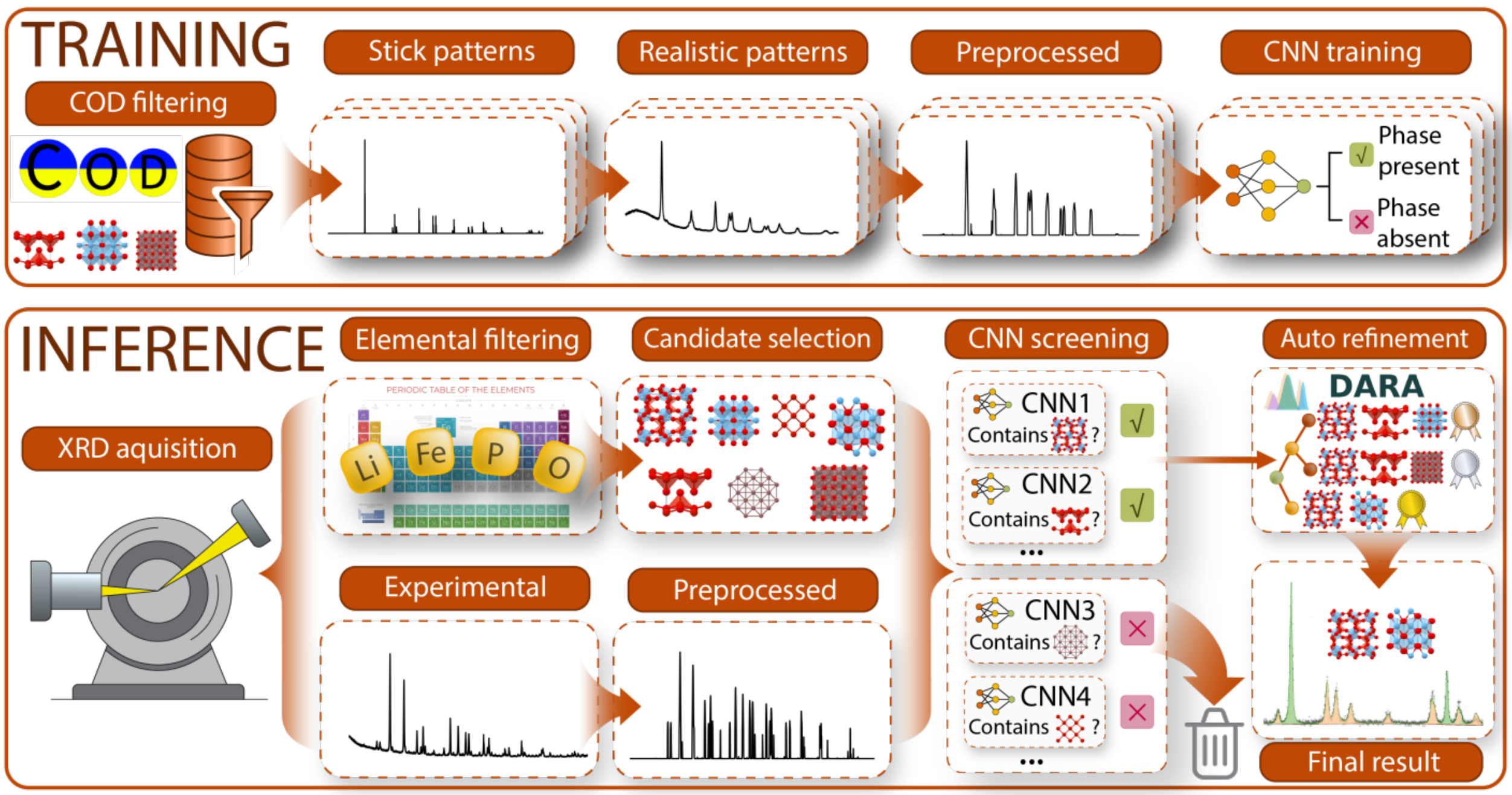

**Figure 1. Overview of the GALAXI workflow.** Top: Crystal structures curated from the COD are used to generate XRD patterns augmented with physical and instrumental artifacts, then preprocessed to train one independent CNN per candidate phase. Bottom: The known chemical space defines the candidate phases, and the supplied XRD pattern is preprocessed identically and screened by the corresponding phase-specific CNNs. Positive candidates are passed to DARA, [16] which performs Rietveld refinement to determine final phase assignments and weight fractions.

**Test Cases**

To evaluate GALAXI, we use a held-out set of 130 experimental patterns spanning 15 phases. Fourteen phases lie within the Fe-Mn-Ti-Li-P-C-O chemical space, while $MoO_3$ is included for a preferred-orientation test. Of the 130 patterns, 61 are from pristine (*i.e.*, not modified in any way) measurements on both single-phase samples and multiphase mixtures. These patterns are used to benchmark GALAXI directly against four baseline methods.[7–10] The remaining 69 patterns form a separate stress-test set designed to probe specific experimental artifacts, including small crystallite size induced by ball milling, uniform peak shifts, sample displacement, and preferred orientation (*i.e.*, texture). These GALAXI-only tests deliberately extend beyond the training distribution to probe the limits of the phase-specific models.

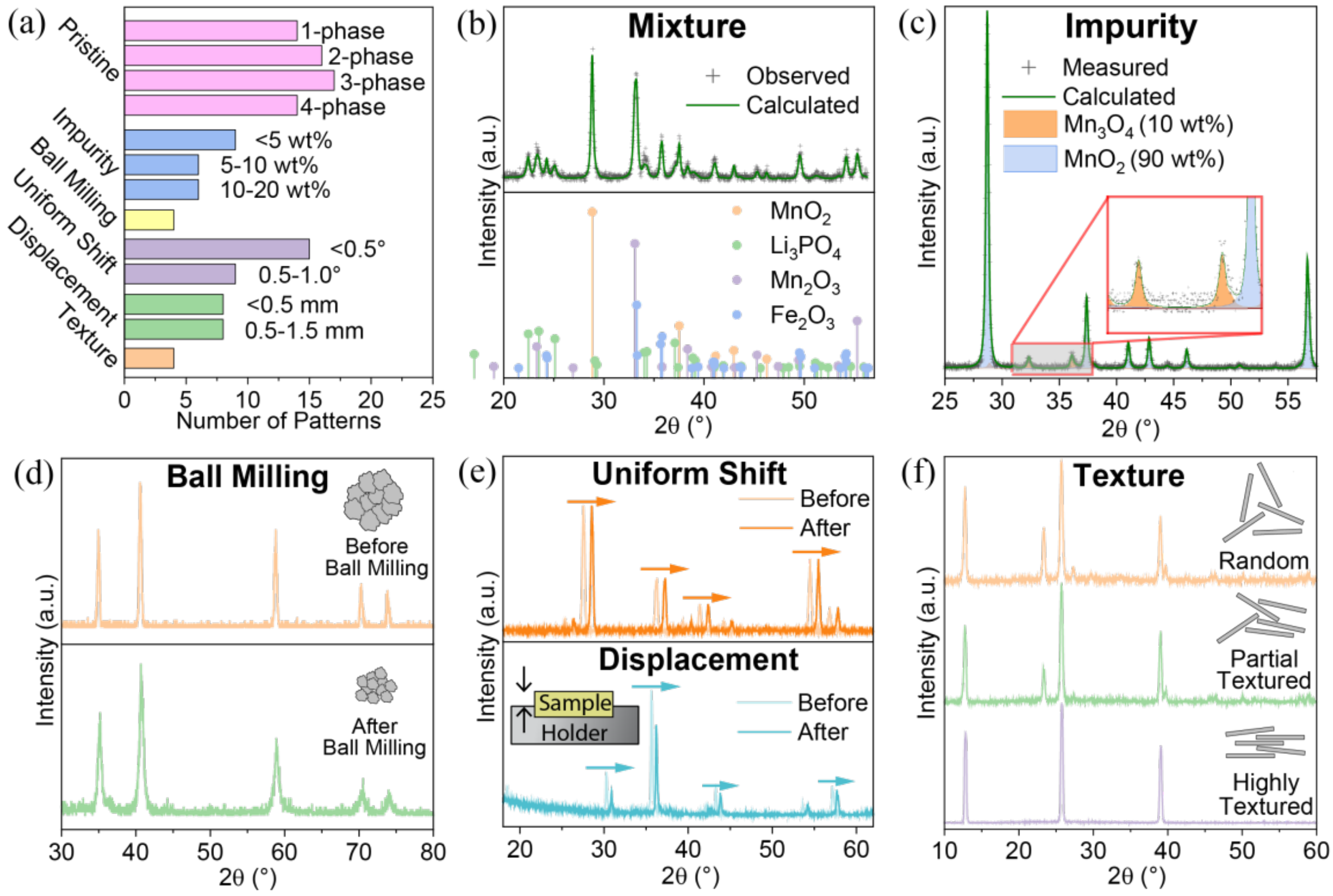

**Figure 2. Overview of the experimental test collection.** (a) Number of XRD patterns in each of the six categories. (b) An example of a multiphase mixture containing $MnO_2$, $Li_3PO_4$, $Mn_2O_3$, and $Fe_2O_3$. Rietveld refinement from DARA is also shown. (c) A secondary phase $Mn_3O_4$ (10 wt%) in $MnO_2$ (90 wt%). (d) Peak broadening before and after ball milling of MnO. (e) Uniform 2θ offset for $TiO_2$ (rutile) and an off-axis sample displacement for $Fe_2O_3$, each shown before and after the shift occurs. (f) $MoO_3$ patterns spanning random, partially textured, and highly textured powders.

**Figure 2a** summarizes the number of experimental patterns in each test category, while panels **b–f** show representative examples of measured data. The pristine set includes both single-phase and multiphase samples, but no deliberate artifacts besides the measurement noise that results from short scan times. This dataset provides the baseline on which we compare the performance of GALAXI against four prior methods. The remaining categories isolate specific experimental difficulties: low phase fraction tests sensitivity to weak minority-phase signals; ball milling introduces peak broadening through reduced crystallite size and microstrain; uniform peak shift and sample displacement probe robustness to systematic peak-position errors; and texture tests sensitivity to preferred orientation using $MoO_3$ as a case study. Details for sample preparation can be found in Methods Section 8.

Beyond the controlled *ex-situ* tests above, we evaluate GALAXI on three *in-situ* XRD datasets collected during high-temperature solid-state synthesis reactions. These provide a more realistic and challenging test because several sources of complexity occur simultaneously, including lattice strain, thermal effects, sample displacement, and transient low-abundance phases. We focus on $BiFeO_3$ formation in the main text, with additional $CaTiO_3$ and $CaCu_3Ti_4O_{12}$ reactions reported in the Supplementary Information.

**Results**

**Pristine samples**

GALAXI naturally produces threshold-based phase assignments since each phase-specific CNN independently reports a probability that a certain phase is present. Using a fixed probability threshold of 0.5, the CNN screening stage achieves a micro-F1 score of 0.304 on the pristine test set, with high recall (0.975) but low precision (0.180). This low micro-F1 reflects the deliberate emphasis on recall, since missing a true phase at this stage is more consequential than retaining extra false positives (*i.e.*, phases that are not present). After passing the CNN-filtered candidates to DARA, micro-F1 increases to 0.935, with recall of 0.929 and precision of 0.941. Performance also remains strong as the number of phases increases, with the full GALAXI pipeline achieving micro-F1 scores of 0.963, 0.941, 0.941, and 0.875 for patterns containing one, two, three, and four phases, respectively (Supplementary Figure S2).

We next compare GALAXI against four baseline methods: XCA,[9] XQueryer,[10] Peak Search-Match and XRD-AutoAnalyzer[7,8]. Implementation details are provided in Methods Section 7. Because these methods return ranked candidate lists without directly comparable probability thresholds, we first compare their raw identification performance (before any DARA refinement) using top-$k$ accuracy. With this metric, a prediction is counted as correct if every phase present in the experimental sample appears within the model's top-$k$ candidates (Methods Section 5). As shown in **Figure 3a**, GALAXI outperforms the baselines from $k \approx 10$ onward, reaching 67.2% top-$k$ accuracy at $k \approx 10$, compared with 39.3% for XCA, the strongest baseline, and 24.6–32.8% for the remaining methods. At $k \approx 15$, GALAXI reaches an accuracy of 82.0% while the baselines give lower accuracies ranging from 32.8 to 47.5%.

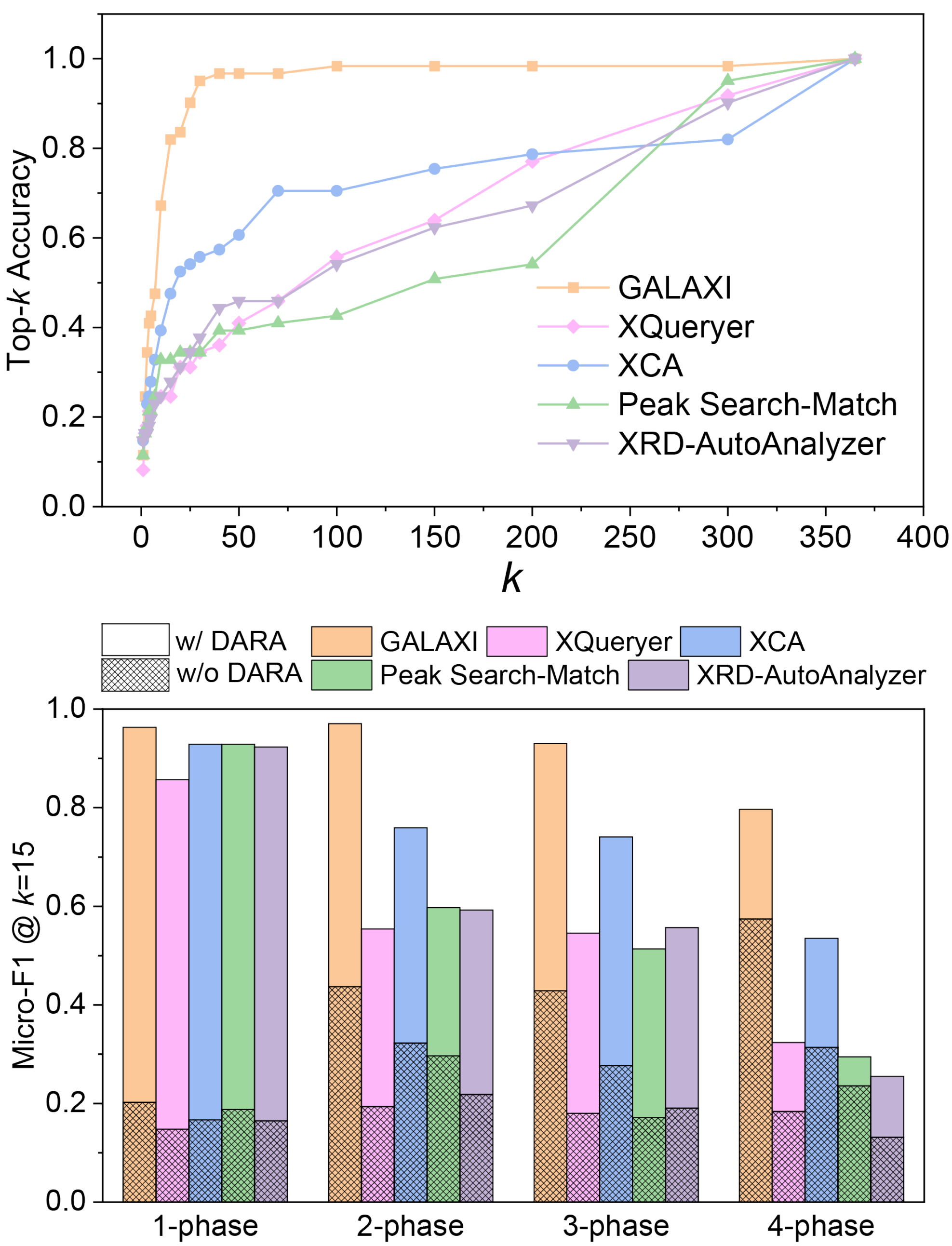

**Figure 3. Cross-model comparison on the pristine 61-pattern test set.** (a) Top-*k* accuracy for GALAXI versus four baselines as a function of *k*. These baselines include XQueryer,[10] XCA,[9] Peak Search-Match, and XRD-AutoAnalyzer[7,8]. (b) Micro-F1 score at *k* = 15, with and without refinement of candidate phases by DARA. Results are broken down by the true number of phases in each pattern.

We next pass each model's top 15 candidate phases to DARA and evaluate the refined predictions as a function of the true number of phases in each pattern (**Figure 3b**). All five models perform comparably well on single-phase patterns, with micro-F1 scores ranging from 0.857 to 0.963. However, GALAXI maintains better performance as mixture complexity increases. It achieves micro-F1 scores of 0.971 at two phases, 0.930 at three phases, and 0.797 at four phases. In contrast, micro-F1 scores from the baselines span a lower range of 0.255 to 0.535 when dealing with four phases in a sample. For GALAXI, these top-15 scores are slightly below its native threshold-based performance because the top-*k* formulation is used only to place all five methods on comparable footing and may either admit low-confidence candidates or exclude a true phase ranked outside the top 15. Using its native threshold-based screening, GALAXI instead reaches a much higher micro-F1 score of 0.875 on four-phase mixtures.

Despite GALAXI's strong performance on the pristine dataset, two related failure modes are worth noting. First, in the most challenging four-phase mixture we tested ($MnO_2$, $LiFePO_4$, $LiMn_2O_4$, and $TiO_2$), the CNN assigned near-zero probabilities to $MnO_2$ and $LiFePO_4$ before DARA could even consider them. Their diagnostic peaks were heavily obscured by other phases: $LiFePO_4$ reflections were strongly suppressed, while $MnO_2$'s major peak near 28° was nearly undetectable (Supplementary Figure S8). Second, DARA can occasionally remove a correctly identified phase or retain a spurious "lookalike" candidate. In a four-phase mixture of $Fe_2O_3$, $Fe_3O_4$, $Li_3PO_4$, and $LiFePO_4$, the CNN identifies $Fe_3O_4$ with high confidence (probability 0.999), but DARA removes it because its reflections overlap with those of $Fe_2O_3$ (Supplementary Figure S9). Conversely, for a single-phase $Li_2TiO_3$ pattern, the CNNs assign high probability to both $Li_2TiO_3$ (0.999) and the structurally similar $LiTiO_2$ (0.999), and DARA retains both because including $LiTiO_2$ gives a better profile fit than just $Li_2TiO_3$ alone (Supplementary Figure S10). This is a general limitation of Rietveld refinement, where adding components can improve the fit even when those components are not physically present. Pre-filtering candidates with CNNs reduces this problem but does not eliminate it.

**Impurities and artifacts**

We next use a series of diagnostic tests to probe how GALAXI's phase-specific CNNs respond to individual experimental artifacts (Methods Section 8). **Figure 4** tracks the CNN probability assigned to the correct phase as each perturbation is increased, including low impurity phase fraction, uniform peak shift, and sample displacement. These controlled tests isolate where the CNN screening stage begins to lose sensitivity as the experimental pattern moves beyond the training distribution.

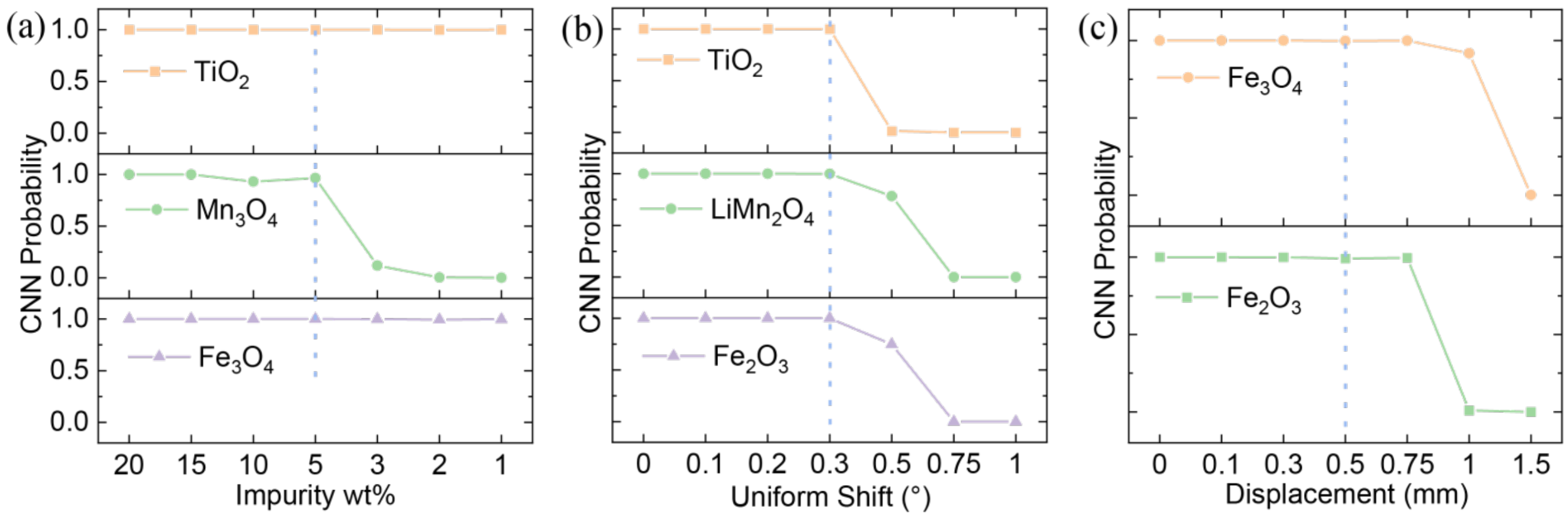

**Figure 4. CNN probability for the correct phase as a function of artifact magnitude.** (a) Probability for $TiO_2$, $Mn_3O_4$, and $Fe_2O_3$ as each phase is diluted from 20 down to 1 wt% in its corresponding two-phase mixture. (b) Probability for $TiO_2$, $LiMn_2O_4$, and $Fe_2O_3$ as uniform 2θ offset increases from 0° to 1.0°. (c) Probability for $Fe_2O_3$ and $Fe_3O_4$ as the sample is deliberately offset from the diffractometer's focusing circle from 0 to 1.5 mm. In all plots, the dashed vertical line represents the limit used during synthetic pattern generation for model training.

On the impurity series, three pairs of phases were tested: $TiO_2$ (rutile)/$TiO_2$ (anatase), $MnO_2$/$Mn_3O_4$, and $Fe_2O_3$/$Fe_3O_4$. In each pair, the latter serves as the "impurity", and its weight fraction is diluted across seven levels from 20 down to 1 wt%. As shown in **Figure 4a**, $TiO_2$ (anatase) and $Fe_3O_4$ remain confidently detected by GALAXI across the entire 20-to-1 wt% range, while $Mn_3O_4$ is the only phase whose probability degrades as its weight fraction drops to 3 wt% and below. It is notable that GALAXI often remains sensitive down to weight fractions $< 5$ wt%, since this is the lower bound of weight fractions that were simulated during training, and the model robustness to such low values suggests at least some ability to generalize.

Single-phase test patterns were shifted by 0.1° to 1.0° in **Figure 4b** to probe robustness against systematic peak-position error. Raw CNN probability for the correct phase remains above 0.999 through a 0.3° shift for all three phases tested ($Fe_2O_3$, $LiMn_2O_4$, $TiO_2$), consistent with the ±0.3° uniform-shift range used during training. It then falls sharply at 0.5°, beyond the limit of uniform peak shift used in the training set, and collapses to near zero at larger shifts (0.75–1.0°).

To probe the effects of peak shifts that are non-uniform, we next performed measurements with sample displacement ($z$). Single-phase $Fe_2O_3$ and $Fe_3O_4$ patterns were re-measured with the sample deliberately displaced from the diffractometer's focusing circle by $z$ = 0 to 1.5 mm (Methods Section 8.3). As indicated in **Figure 4c**, the raw CNN probability for both phases remains > 0.99 up to a displacement of 0.75 mm, extending beyond the ±0.5 mm displacement range used during training. The probability for $Fe_2O_3$ collapses to 0.009 by $z$ = 1.0 mm, while the probability for $Fe_3O_4$ degrades more gradually to 0.917 by the same value. Both phases have near-zero probability by 1.5 mm displacement.

On a small set of single-phase samples that were subjected to ball milling, which broadens diffraction peaks through crystallite-size reduction and/or microstrain (**Figure 2d**), GALAXI achieves a micro-F1 score of 0.889 using its default probability threshold of 0.5. This artifact is not particularly challenging for GALAXI, likely because the observed broadening is modest relative to the wider 5.5–450 nm crystallite-size range used during training. Performance may deteriorate under more extreme broadening, such as for crystallite sizes approaching or falling below the lower end of the training range or for severe microstrain.

Finally, we assess GALAXI's sensitivity to peak-intensity variations by considering $MoO_3$, a layered compound with a strong intrinsic tendency toward preferred orientation. The as-prepared drop-cast sample is highly textured, as shown in **Figure 2f**, causing its relative peak intensities to deviate substantially from those expected for a randomly oriented powder. We progressively reduce this texture by milling the powder before drop-casting, with longer milling times producing increasingly random crystallite orientations. For the fully textured sample (0 min milling), the CNN assigns $MoO_3$ a low probability of only 0.052. This probability increases sharply to 0.998 after just 1 minute of ball milling and reaches 1.000 after 20 minutes (Supplementary Figure S5). These results show that GALAXI can become unreliable when strong preferred orientation produces intensity distributions that differ substantially from the randomly oriented powder patterns represented during training.

***In-situ* data**

As a final test case, we applied GALAXI to analyze a series of *in-situ* XRD patterns collected during high-temperature solid-state synthesis reactions targeting three separate oxides: $BiFeO_3$, $CaTiO_3$, and $CaCu_3Ti_4O_{12}$. These experiments introduce new artifacts that are distinct from the room-temperature *ex-situ* patterns evaluated above, including lattice strain, sample displacement from the furnace-stage geometry, temperature-dependent Debye-Waller attenuation, and transient, low-weight-fraction intermediate phases characteristic of an evolving reaction pathway.[19–21]

The $BiFeO_3$ solid-state synthesis reaction starts from a stoichiometric mixture of $Bi_2O_3$ and $Fe_2O_3$, with 31 patterns collected in equal time intervals as the sample is heated from 50 to 800 °C (Methods Section 8). The resulting XRD heatmap and phases identified by GALAXI are shown in **Figure 5**. Our models indicate that $Bi_2O_3$ and $Fe_2O_3$ persist up to ~475 °C, after which a Bi-rich intermediate phase forms. This phase is consistent with the sillenite-type $Bi_{25}FeO_{39}$ phase, first appearing at 500 °C and rising sharply to a maximum 70.9% weight fraction at 600 °C before declining as the targeted $BiFeO_3$ product forms and becomes the dominant phase, with a weight fraction rising from 11.8% at 625 °C to 47.8% by 800 °C.

These results are in good agreement with the reaction pathway independently reported in the literature for $BiFeO_3$, which forms through the $Bi_{25}FeO_{39}$ intermediate rather than directly from the binary oxide precursors.[22] GALAXI also identifies the known secondary phase $Bi_2Fe_4O_9$ (mullite-type), which emerges alongside $BiFeO_3$ from 725 °C onward, consistent with the incomplete synthesis reactions often reported in the literature.[23] However, performance is not perfect across the entire dataset: at 550 °C, $Fe_2O_3$'s refined weight fraction briefly drops to zero even though the CNN continues to confidently flag its presence (probability ~0.998). This is a transient artifact of severe peak overlap between $Fe_2O_3$ and the emerging $Bi_{25}FeO_{39}$ phase at that temperature, which causes DARA to exclude it entirely.

We also applied GALAXI to the reactions: $CaCO_3 + TiO_2 \rightarrow CaTiO_3 + CO_2$ (37 patterns, 50–950 °C) and $CaCO_3 + 3\ CuO + 4\ TiO_2 \rightarrow CaCu_3Ti_4O_{12} + CO_2$ reaction (43 patterns, 50–1100 °C). The results are summarized in Supplementary Note S4. The general pathways in both reactions are correctly resolved, including the detection of a short-lived CaO decomposition product for the first synthesis reaction, and small amounts of the $CaTiO_3$/$Ca_4Ti_3O_{10}$ intermediates formed during the latter reaction, consistent with pathways reported in the literature.[24]

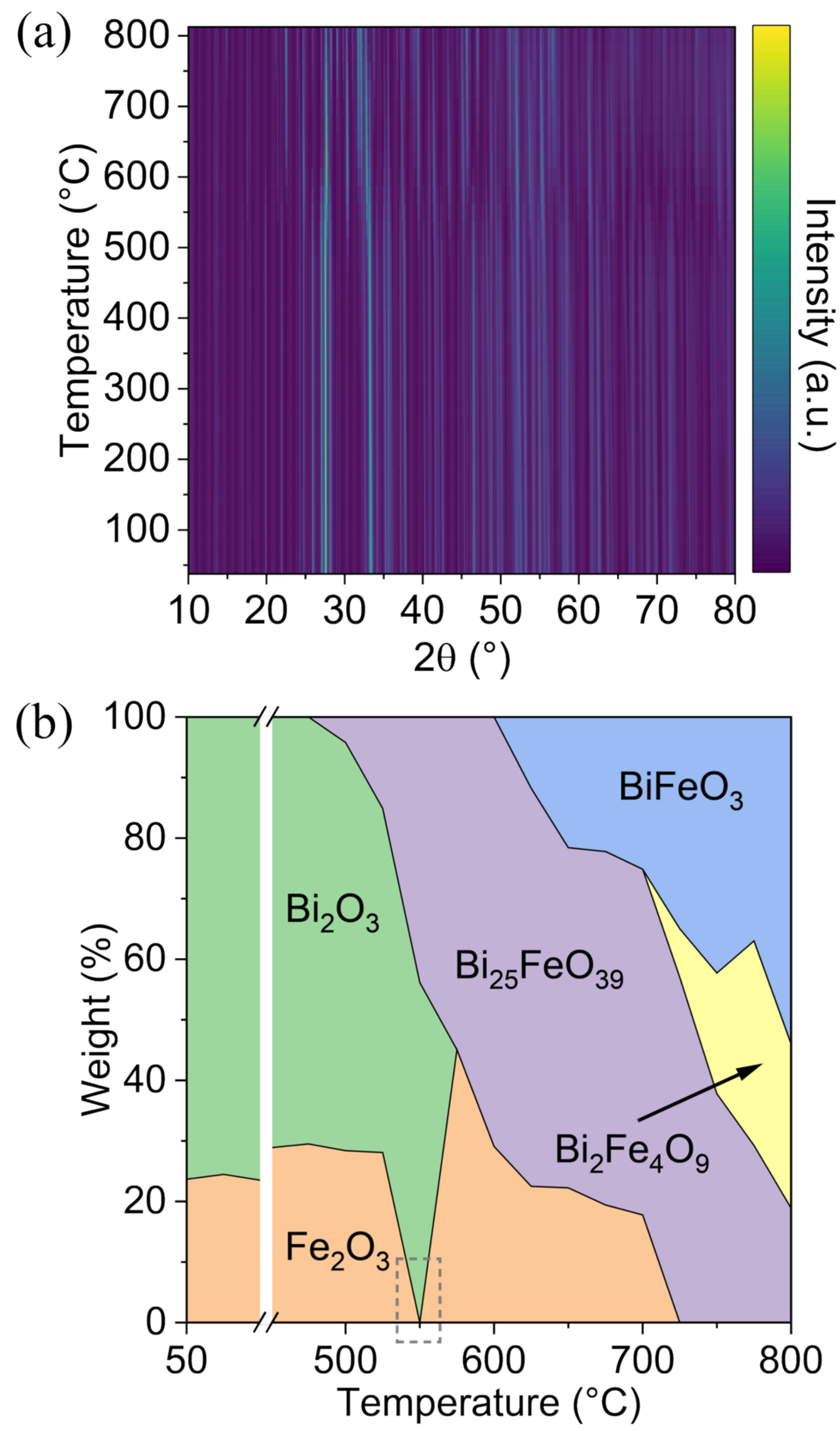

**Figure 5. *In-situ* XRD of the $Bi_2O_3$ + $Fe_2O_3$ → $BiFeO_3$ solid-state synthesis reaction**. (a) Raw X-ray diffraction intensity plotted as a function of temperature. (b) GALAXI-identified phases with DARA-refined weight fractions as a function of temperature. The gray square highlights the abnormal point where the refined weight fraction of $Fe_2O_3$ briefly drops to zero.

**Scaling with Reference Pool Size**

A key strength of GALAXI's architecture is that its library of phase-specific models grows independently of any single chemical space. Because each candidate phase is served by its own dedicated model, new phases can be added without retraining the full existing library. This also allows the CNNs to serve as an efficient prescreening step before combination-level refinement by DARA, reducing the number of candidate phases that DARA must consider. This is important because DARA evaluates combinations of candidate phases, so its computational cost can grow rapidly with the size of the reference pool. To quantify this benefit, we examine how model performance and computational cost scale as the number of candidate phases increases.

As shown in **Figure 6**, we vary the number of candidate phases from 21 (the minimum phases required to cover the pristine test set) to the full 365-phase reference catalog of phases spanning the Fe-Mn-Ti-Li-P-C-O chemical space. At each catalog size, the same test patterns are evaluated using three workflows: CNN-only screening, CNNs + DARA, and DARA-only refinement. With CNNs + DARA, our models filter the reference catalog and only positively identified candidates with high CNN output probability are passed to DARA. In the DARA-only workflow, the entire reference catalog at that size is passed directly to DARA without CNN prescreening.

The combination of CNNs and DARA maintains a high and relatively stable micro-F1 score, decreasing only slightly from 0.947 at 21 phases to 0.935 at 365 phases. Importantly, CNN prescreening keeps the number of phases passed to DARA small even as the full reference pool expands (Supplementary Figure S12), limiting the combinatorial search during refinement. As a result, the total computational cost increases only modestly, from 5.1 to 24.5 s per pattern. In contrast, without CNN prescreening, DARA must search directly over the expanding reference pool, causing its computational cost to increase from 15.4 to 862.4 s per pattern. This larger search space is also accompanied by a decline in DARA-only performance, with micro-F1 decreasing from 0.929 to 0.817. The CNN-only workflow shows a different limitation: its micro-F1 drops sharply to 0.304 at 365 phases as the expanding reference pool introduces more false-positive candidates. The precision and recall in Supplementary Figure S11 clarify this trend: CNN-only recall remains nearly constant at 0.975, whereas precision falls from 0.560 to 0.180. In contrast, CNNs + DARA maintain high precision (0.941–0.966) and recall (0.928–0.929), demonstrating the practical value of CNN prescreening for scaling DARA to large reference pools.

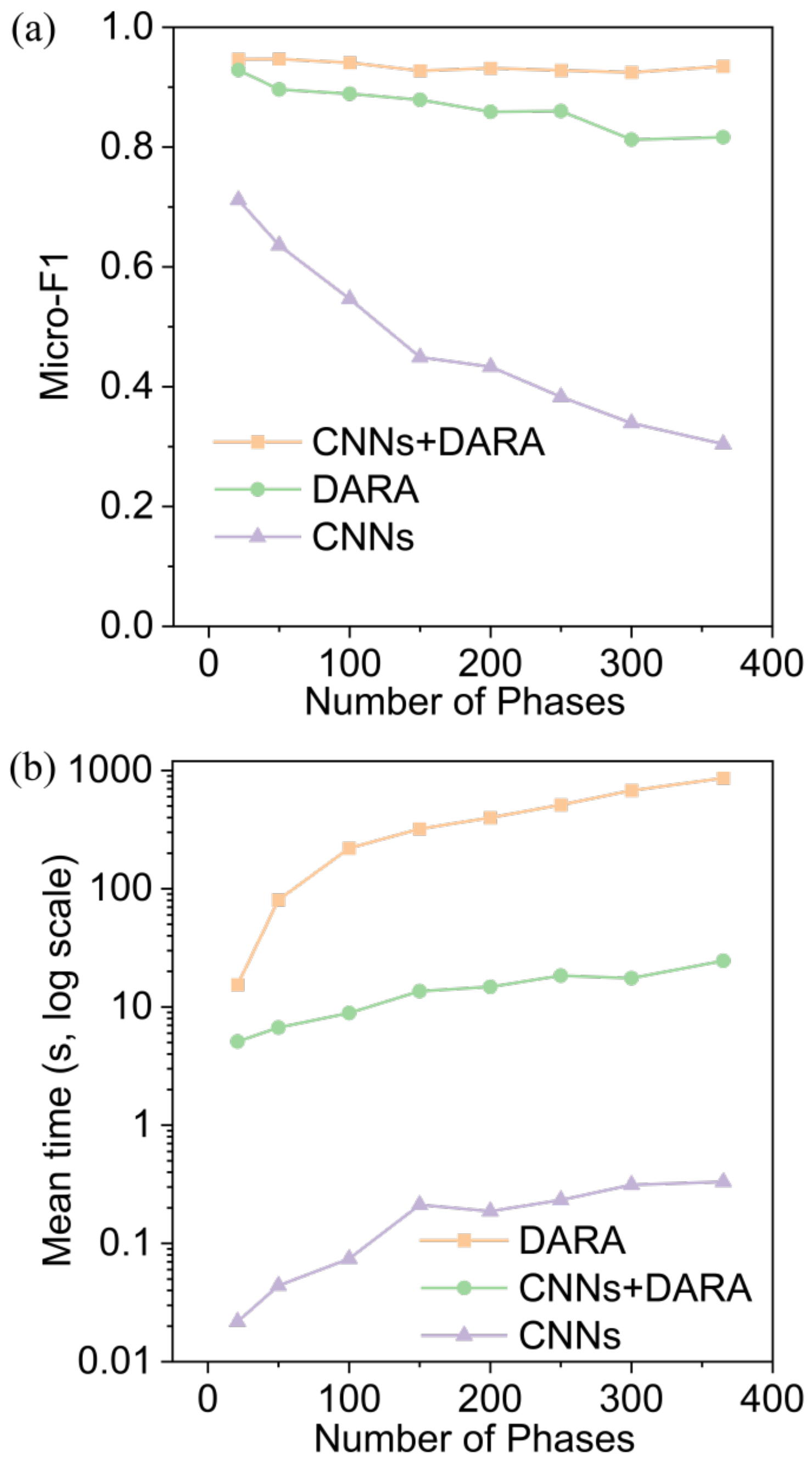

**Figure 6. Scaling of performance and computational cost with reference-catalog size.** (a) Micro-F1 score as a function of the number of candidate phases for CNN-only, CNNs + DARA, and DARA-only phase identification on pristine experimental XRD patterns containing one to four phases. (b) Mean inference time per pattern (log scale) for the same three workflows.

## Public models trained on the COD

Having established that GALAXI can scale to large reference catalogs without retraining existing models, we next apply this architecture across the Crystallography Open Database (COD).[18] After filtering and deduplicating the database for physical plausibility and structural redundancy (Methods Section 1), we train phase-specific classifiers for 64,594 structures. This greatly expands the chemical and structural space accessible to GALAXI beyond the systems used for experimental benchmarking. **Figure 7** summarizes the elemental, compositional, and structural coverage of this pretrained library.

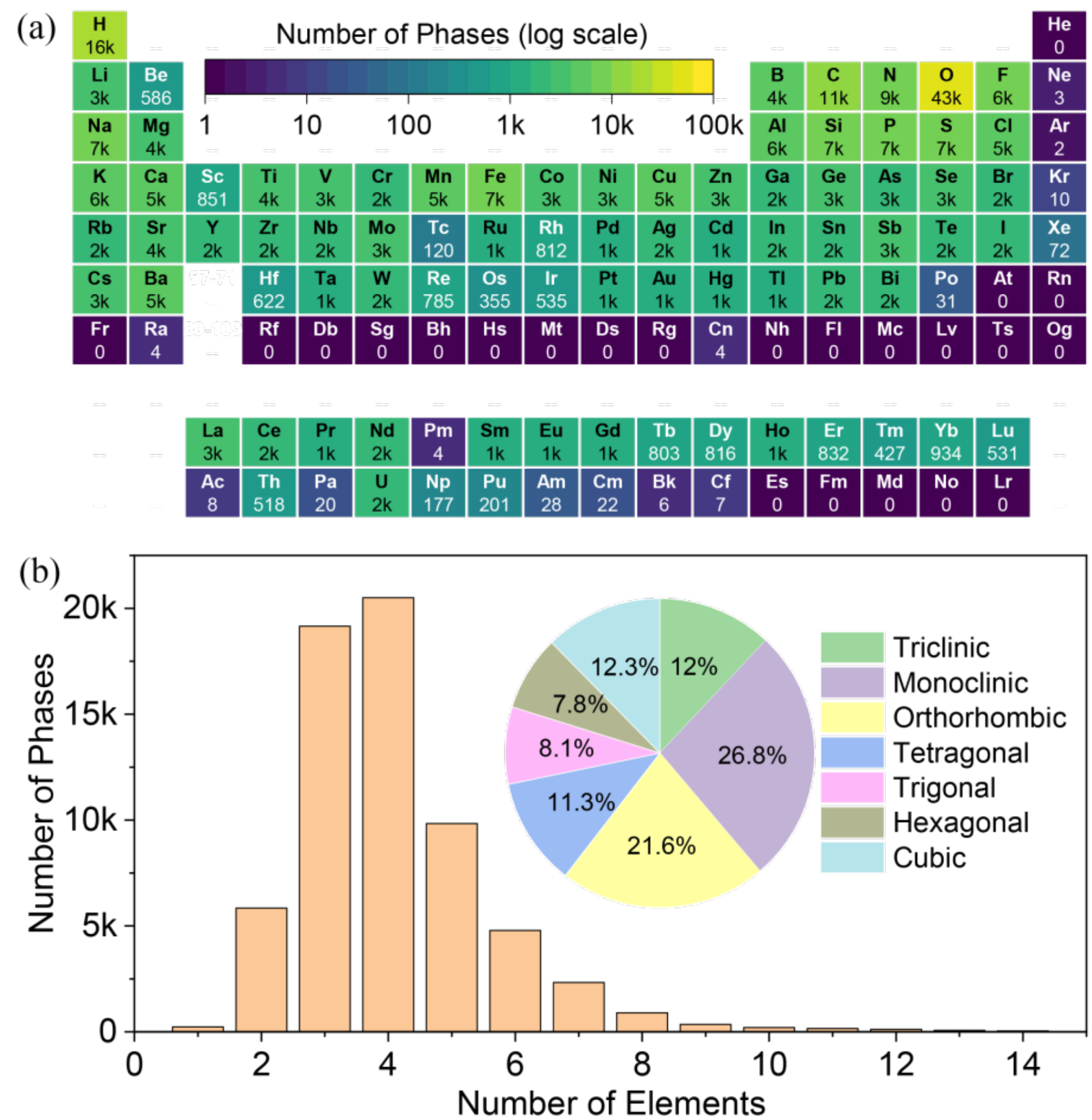

**Figure 7. Coverage of the pretrained model library spanning 64,594 phases**. (a) Periodic table heatmap of the number of pretrained models per element. (b) Distribution of the number of distinct elements per phase. Inset: distribution of crystal systems.

The periodic table heatmap (**Figure 7a**) shows broad chemical coverage. Oxygen is the most represented element, appearing in 42,770 phases, followed by hydrogen (16,218) and carbon (11,168), with extensive representation across the transition and post-transition metals. The library is dominated by ternary and quaternary compounds (**Figure 7b**), with a long tail extending to phases containing up to 21 distinct elements. Monoclinic (26.8%) and orthorhombic (21.6%) structures are the most common, while hexagonal structures are least represented (7.8%). Additional statistics on unit-cell volume, volume per atom, and density are provided in Supplementary Figures S14–S16.

Each phase-specific model is relatively lightweight, containing approximately 0.46 million trainable parameters and occupying 1.83 MB of disk space. Training-data generation and model training require an average of 47.0 s and 16.8 s per phase on a GPU, respectively. Most models (81%) train within 20 s and nearly all within 50 s, while training-data generation shows a longer tail for structurally complex phases (Supplementary Figure S13).

This pretrained library is deployed through a public web interface at https://galaxi-xrd.com, where users can upload XRD patterns, define the relevant chemical space, screen candidate phases using the pretrained CNN library, and inspect the resulting CNN probabilities alongside DARA-refined phase assignments and quantitative phase fractions.

**Discussion**

Taken together, the results presented here suggest that decoupling phase detection from combination-level refinement is a practical strategy for multiphase XRD identification. The phase-specific CNNs used by GALAXI can be optimized for detection of individual phases without requiring a single model to discriminate among an ever-growing reference set, while DARA provides the downstream refinement needed to determine which combination of candidates best explains the diffraction pattern. This division of labor is especially useful for complex mixtures, where weak or overlapping phase signatures are difficult to resolve, and allows the reference library to expand without retraining existing models. The experimental artifact tests and *in-situ* synthesis examples further indicate that this framework remains useful beyond idealized single-phase measurements, although important limitations remain.

The first limitation is that GALAXI's independent phase-specific detectors, while robust to most artifacts tested, can still miss a true phase in sufficiently complex mixtures. The downstream DARA refinement can also introduce errors by either dropping a correctly identified phase when its peaks strongly overlap with another phase or retaining a spurious, structurally similar phase alongside a correct identification. Both failure modes reflect a trade-off introduced by DARA's role in the pipeline: because our modified DARA implementation does not penalize peak overlap between candidate phases (Methods Section 4), a larger or less-curated starting candidate pool provides more opportunities for incorrect combinations to be retained. This behavior is a consequence of the modification used here to keep the combinatorial search tractable.

A second limitation is that GALAXI's performance degrades sharply for heavily textured samples, since the synthetic training patterns assume a statistically random powder average and therefore do not capture strong preferred-orientation effects. This may limit performance in materials-characterization settings where texture is common, such as thin films and anisotropic nanocrystals.[25] In principle, this limitation could be addressed by expanding the range of texture-related artifacts represented during training. However, doing so may reduce accuracy for conventional powder samples, highlighting that the optimal training distribution depends on the intended application and should reflect the experimental conditions of interest.

A final limitation is GALAXI's sensitivity to systematic peak-position errors, particularly uniform peak shifts and sample displacement. This again reflects a deliberate choice in the training distribution rather than a fundamental limitation of the architecture. The ranges used for these artifacts were chosen to balance robustness to moderate peak shifts against the increased false-positive rate that can arise when larger shifts cause reflections from different phases to overlap.

Despite these limitations, GALAXI performs well across the broad range of experimental conditions considered here. Beyond predictive accuracy, a key advantage of the framework is its scalability. Because individual phase models can be trained and added independently, the reference library can continue to expand as new structures become available without retraining existing models. The pretrained models developed in this work are made publicly accessible through https://galaxi-xrd.com, providing a practical foundation for broader application and continued extension to new materials systems.

**Acknowledgements**

We acknowledge computational resources provided by the Delta supercomputer through the NSF ACCESS program and by the Hoffman2 Shared Cluster at UCLA. Additional computational resources were supported by the NVIDIA Academic Grant Program. This work was also supported by startup funds from the UCLA Samueli School of Engineering.

**Author Contributions**

X. Tong: model architecture design and tuning, preparation of experimental test set samples, and drafting of the manuscript. E. Jin: design and testing of the public website. J. Xu and A. Rao: preparation of experimental test set samples. P. Jiang: model architecture and training design; N. J. Szymanski: conceptualization, daily supervision and mentoring, and review and editing of the manuscript.

**Code and Data Availability**

All code used in this manuscript is available at https://github.com/Szymanski-Group/galaxi. This repository also includes the experimental XRD patterns used for testing. Pre-trained CNNs from GALAXI are available at https://doi.org/10.6084/m9.figshare.33360183. Baseline models, used for comparison, can be found at: https://doi.org/10.6084/m9.figshare.33420550 (XQueryer), https://doi.org/10.6084/m9.figshare.33420511 (Crystallography Companion Agent, XCA), and https://doi.org/10.6084/m9.figshare.33420508 (XRD-AutoAnalyzer).

## Methods

### 1. Dataset curation

#### 1.1 Data filtering

Raw Crystallographic Information Files (CIFs) were first parsed and subjected to multiple filters. We excluded unit cells containing more than 200 atoms since these require long compute times for the generation of their XRD patterns. We also excluded structures composed entirely of non-metallic elements (such as $CO_2$ and $O_2$) to instead focus on materials that are solid under ambient conditions. Further extending our focus on ambient conditions, we excluded structures that were measured outside the 100–400 K range or at pressures exceeding 0.1 GPa. When multiple entries were present for the same materials, we prioritized those with more recent measurements (published in or after 1990), falling back on older entries only when necessary.

**1.2 Data grouping**

In some cases, the COD contains multiple closely related structural entries, such as slightly off-stoichiometric variants, that produce effectively indistinguishable XRD patterns. Because distinguishing between such entries is beyond the scope of GALAXI and, in many cases, powder XRD itself, we group structurally equivalent entries before evaluation. Clustering was performed using the StructureMatcher functionality in pymatgen,[26] with fractional length, site, and angle tolerances of 0.2, 0.5, and 10.0°, respectively, together with a partial-occupancy tolerance of 0.2. For each resulting cluster, we selected the medoid as the representative structure, defined as the member with the lowest average root-mean-square displacement relative to all other members. Structures with a minimum interatomic distance below 1.0 Å or an atomic number density outside the range of 0.01–0.15 atoms/$Å^3$ were excluded from the final dataset.

**2. Pattern Generation**

**2.1 Modeling of Physical and Instrumental Artifacts**

To bridge the domain gap between idealized XRD patterns and real-world experimental measurements, we developed a modeling approach that applies physical and instrumental artifacts to the synthetic data. Exact sampled ranges for each parameter below, together with their effect on validation performance, are given in the Supplementary Table S1.

**2.1.1 Lattice strain**

Macro-scale homogeneous lattice strain ($\varepsilon$), which can arise from chemical substitution, thermal expansion, or residual stress, is explicitly applied to the unit cell before diffraction profile generation to introduce systematic peak-shifting artifacts. The transformation of the crystal unit cell is governed by:

$$\boldsymbol{F} = \boldsymbol{I} + \varepsilon$$

where $\boldsymbol{I}$ is the $3 \times 3$ identity matrix and $\varepsilon$ is the symmetric strain tensor containing components sampled randomly within predefined boundaries $\varepsilon_{ij} \in [-s_{\min}, s_{\max}]$. To prevent breaking of symmetry, the sampling of $\varepsilon$ is bound by crystal system constraints. For example, $\varepsilon_1 = \varepsilon_2 = \varepsilon_3$ is enforced for cubic materials. The original crystallographic lattice matrix $\boldsymbol{L_0}$ is transformed into a strained variant *via* matrix multiplication:

$$\boldsymbol{L}_{\mathbf{strained}} = \boldsymbol{L_0}\boldsymbol{F}$$

All peak positions in XRD are then computed *via* the modified interplanar spacings $d_{hkl}$.

### 2.1.2 Thermal effects

Thermal atomic vibrations increasingly attenuate high-angle reflections. This attenuation is explicitly simulated *via* the Debye-Waller factor:[27]

$$I = I_0 \exp(-2M)$$

where the exponent

$$M = 8\pi^2 u_{\text{iso}} (\sin^2\theta / \lambda^2)$$

scales with the isotropic atomic displacement parameter $u_{\text{iso}}$, which is randomly sampled from a reasonable range of values to span the thermal disorder expected across sample compositions and measurement temperatures.

Thermal effects not only affect peak intensity but also affect background, which necessitates modeling of diffuse scattering effects. Thermal diffuse scattering is modeled as:

$$I_{\text{diffuse}} = I'_0 [1 - \exp(-2M)]$$

where $I'_0$ is an intensity scaling factor while $M$ is defined the same as above. $I_{\text{diffuse}}$ is then added to the background.

### 2.1.3 Peak Shapes and Broadening Effects

Incident radiation is modeled as a superposition of the characteristic Cu K$\alpha_1$ ($\lambda_1$=1.5406 Å) and Cu K$\alpha_2$($\lambda_2$ = 1.5444 Å) emission doublet, with theoretical intensity ratio of $I_{\alpha_1}/I_{\alpha_2} \approx 2$. The discrete peak lists are then converted to a profile distributed across the whole diffraction angle 5° to 105° using a pseudo-Voigt profile. This is formulated as a linear combination of Gaussian and Lorentzian components:[28]

$$P(\Delta 2\theta) = (1 - \eta)\exp[-\frac{1}{2}(\frac{\Delta 2\theta}{\sigma})^2] + \eta \frac{\gamma^2}{(\Delta 2\theta)^2 + \gamma^2}$$

where $\Delta 2\theta$ is the angular deviation from the theoretical reflection center, and $\eta$ is a dynamically sampled mixing parameter governing the fractional Lorentzian contribution. The width parameters $\sigma$ and $\gamma$ are mapped from the full-width at half-maximum (FWHM) *via*:

$$\sigma = \text{FWHM}/2\sqrt{2\ln 2} \text{ and } \gamma = \text{FWHM}/2.$$

As broadening is composed of three components, namely instrumental broadening, size broadening and strain broadening, FWHM is thus calculated as:

$$\text{FWHM}_{\text{total}} = \sqrt{\text{FWHM}^2_{\text{instrumental}} + \text{FWHM}^2_{\text{size}} + \text{FWHM}^2_{\text{strain}}}$$

$$\mathrm{FWHM}_{\mathrm{instrumental}} = \sqrt{U\tan^2\theta + V\tan\theta + W}$$

$$\mathrm{FWHM}_{\mathrm{size}} = \frac{0.9\lambda}{D\cos\theta}$$

$$\mathrm{FWHM}_{\mathrm{strain}} = 4\varepsilon\tan\theta$$

These are the Caglioti equation,[29] the Scherrer equation,[30] and the microstrain broadening formula.[31] $U$, $V$ and $W$ are instrumental parameters, $D$ stands for crystallite size, while $\varepsilon$ defines the root-mean-square microstrain. All these parameters are sampled randomly from reasonable ranges of values during pattern generation.

### 2.1.4 Sample displacement

Sample displacement errors can lead to non-uniform peak shift, especially at lower angles. Sample height displacement $z$ (randomly sampled) relative to the goniometer radius $R$, is modeled to induce systematic angular shifts:[32]

$$\Delta 2\theta = -\frac{2z\cos\theta}{R}$$

### 2.1.5 Pattern background

To simulate the diffuse "hump" present in many XRD patterns at low angles caused by air scattering or sample holder, we added a broad Gaussian profile plus a slowly decaying linear ramp, with randomly sampled Gaussian hump center and width, as well as ramp slope:

$$\mathrm{Background}(x) = I_{\max}\beta[\lambda\exp(-\frac{(x-c)^2}{2\sigma^2} + (1-\lambda)sx]$$

where $I_{\max}$ stands for maximum intensity of the original pattern, $\beta$ is the relative intensity of the diffuse background, $\lambda$ is the mixing weight between Gaussian hump and linear ramp, $c$ and $\sigma$ are the center and width of the Gaussian hump and $s$ is the slope of the linear ramp. The background is then added to the pattern profile.

### 2.1.6 Texture

Preferred crystallographic orientation is modeled *via* a March-Dollase correction.[33] For a pattern, one preferred direction is drawn from a fixed set of ten low-index crystallographic directions (the three principal axes, three face diagonals, and four body diagonals); the angle α between this direction and each reflection's scattering vector scales the intensity by

$$\mathrm{P}(\alpha) = (r^2\cos^2\alpha + \sin^2\alpha/r)^{-3/2}$$

where the March coefficient $r$ is sampled randomly from 0.5 to 1.5. $r = 1$ reproduces a random, untextured powder average, while values of $r$ moving away from 1 in either direction increasingly favor or suppress reflections aligned with the pattern's chosen preferred-orientation direction.

### 2.2. High-Throughput Acceleration Strategies

Scaling this workflow to generate training patterns for thousands of phase-specific models creates a substantial computational bottleneck. We therefore implemented several acceleration strategies for synthetic pattern generation and model training. First, diffraction-pattern generation is vectorized in PyTorch using multidimensional tensor operations. Peak positions, intensities, and dynamically sampled broadening parameters for thousands of reflections are processed concurrently, and the corresponding pseudo-Voigt profiles are evaluated across the full 2θ grid in a single batched operation. This replaces iterative peak-by-peak CPU calculations with parallel GPU computation and substantially reduces pattern-generation time.

Second, impurity-phase simulation is decoupled from the real-time generation loop. Calculating diffraction patterns directly from crystal structures requires repeated symmetry analysis and structure-factor calculations, which becomes expensive when impurity phases are sampled for every training pattern. We therefore precompute 50,000 diffraction profiles from impurity structures randomly sampled from the COD and store them in HDF5 format. During training-data generation, impurity profiles can then be sampled directly from this library rather than recalculated from the underlying crystal structures.

Finally, we use in-memory caching to reduce disk input/output (I/O) and maintain high GPU utilization during model training. Rather than repeatedly regenerating or loading the same synthetic patterns from disk, generated batches are retained in memory and reused during training.

## 3. Preprocessing and Model Architecture

### 3.1 Data Preprocessing

#### 3.1.1 Masking and Smoothing

Raw XRD patterns are first bounded by a Boolean mask to isolate the valid angular range and explicitly exclude missing or artifact-prone data points. The isolated valid intensity data undergo a primary smoothing step using a Savitzky-Golay filter[34] (a 21-point window in the frozen configuration) with a second-order polynomial and reflective boundary conditions to attenuate high-frequency noise without substantially degrading peak morphology.

### 3.1.2 Two-Pass Background Subtraction

A two-pass Statistics-sensitive Non-linear Iterative Peak-clipping (SNIP) algorithm[35] is employed for background estimation and removal. The SNIP algorithm isolates the underlying baseline by iteratively comparing each data point to its neighbors. Specifically, the estimated background intensity $z$ at the $i$-th data channel (or $2\theta$ step) and iteration step $p$ is updated according to:

$$z_i^{(p)} = \min(z_i^{(p-1)}, \frac{z_{i-p}^{(p-1)} + z_{i+p}^{(p-1)}}{2})$$

The algorithm initializes $p$ at a predefined maximum iteration window size and then decrements down to 1.

The frozen configuration sets the global pass's maximum iteration window to 24. The initial pass captures the global background profile using a large iteration window size. Subtraction of this global baseline yields a partially leveled signal, which is subjected to a subsequent, highly localized SNIP pass with a smaller iteration window size. Subtraction of this local background extracts the raw diffraction peak residuals.

### 3.1.3 Noise Gating and Non-linear Magnification

CNNs are generally less sensitive to weak diffraction peaks, particularly when they occur alongside much stronger reflections. We therefore magnify low-intensity signal to improve the visibility of these features before classification. However, before magnifying weak signal, it is necessary to isolate true diffraction peaks from residual background artifacts, otherwise noise would also be amplified. Accordingly, our algorithm reduces this noise in two distinct ways: by applying a static baseline offset and by defining a dynamic variance threshold.

First, a static noise floor is established at the $10^{th}$ percentile of the residual signal. This floor is uniformly subtracted from the pattern to enforce a strict zero-minimum, and any resulting negative intensities are clamped to zero. Concurrently, the amplitude of noise fluctuation is quantified using the Median Absolute Deviation (MAD),[36] scaled to approximate the standard deviation of the noise $\sigma$:

$$\sigma = 1.4826 \times \mathrm{median}(|x - \mathrm{median}(x)|)$$

where $x$ represents the residual intensities.

Following baseline subtraction, the cleaned signal is max-normalized to a $[0, 1]$ scale ($y_{\mathrm{norm}}$). To selectively amplify low-intensity minor phases while actively suppressing the remaining background noise, a non-linear exponential treatment is applied. The estimated noise spread $\sigma$ is scaled by noise sensitivity $s$ to define a dynamic gating threshold $\sigma_{dyn} = \sigma \cdot s$, which dictates the boundary between noise and true signal. The final magnified intensity $y_{mag}$ is calculated as

$$y_{\mathrm{mag}} = (y_{\mathrm{norm}}[1 - \exp(-(\frac{y_{\mathrm{norm}}}{\sigma_{\mathrm{dyn}}})^{\gamma})])^{\beta}$$

where $\gamma$ dictates the sharpness of the noise-gating threshold and $\beta$ $(0 < \beta < 1)$ controls the magnification of the peaks. Our frozen configuration sets noise sensitivity to 6.0, gate sharpness to 6.0, and the magnification exponent to 0.3.

### 3.2 CNN architecture

The network input is a 1D preprocessed XRD intensity profile spanning at least 20°–60° in 2θ and up to 5°–105°. Profiles extending beyond this range are cropped. Each pattern is resampled onto a common angular grid containing 7001 points, with regions outside the experimentally measured range padded to produce a fixed-length input. A corresponding 7001-dimensional binary mask is provided as a second input channel, with 1 indicating experimentally measured regions and 0 indicating padded regions.

The feature extractor comprises three sequential Conv1d–BatchNorm1d–ReLU blocks with 8 channels and kernel sizes of 27, 15, and 11, followed by max-pooling factors of 2, 2, and 1, respectively (Supplementary Figure S17). A mask-modulated spatial attention module is applied to the final feature maps. A 1 × 1 convolution generates an attention score at each angular position, while the input mask is downsampled to match the feature-map resolution. Positions outside the valid experimental range are assigned an attention score of negative infinity before applying a spatial softmax, ensuring that padded regions receive zero attention. The resulting attention weights are multiplied element-wise with the feature maps, suppressing boundary artifacts and focusing the network on valid diffraction features. The weighted feature maps are flattened and passed to an 8-dimensional fully connected layer followed by a single sigmoid output neuron, which gives the predicted probability that the corresponding phase is present.

### 3.3 Model training

The synthetic data were stratified into training, validation, and testing subsets following a 90/5/5 percentage allocation. These data are discussed in Supplementary Note S2. The network optimization process was conducted using a batch size of 32 patterns. The training objective was formulated around the minimization of the Binary Cross-Entropy (BCE) loss function to optimize the binary classification performance of the network. The optimizer was RMSprop. An early stopping patience of 5 was set and the model checkpoints were monitored continuously against the validation loss to safeguard the network against overfitting during optimization.

To help orchestrate model training, the Weights & Biases (wandb) framework was integrated. Training routines were parallelized as asynchronous workers distributed across available compute resources, with each worker independently optimizing a model for a specific phase classification task. Training was performed on one NVIDIA RTX PRO 6000 Blackwell Max-Q Workstation Edition GPU, with additional model training and hyperparameter tuning carried out on the Delta system at the National Center for Supercomputing Applications, as supported by the Advanced Cyberinfrastructure Coordination Ecosystem: Services & Support (ACCESS) program.

## 4. Rietveld Refinement

Rietveld refinement was performed using a modified version of DARA, a tree search-based framework for automated phase identification from powder X-ray diffraction data.[16] DARA constructs a search tree in which each node represents a candidate phase combination together with its Rietveld refinement result. At each level of the tree, candidate phases are first ranked using a peak-matching score and then individually added to the current phase assembly for full-pattern Rietveld refinement. The refinement results are subsequently evaluated to determine which branches should be further expanded, allowing the algorithm to efficiently explore combinations of candidate phases while avoiding exhaustive enumeration.

In this work, several modifications were introduced to improve both the computational efficiency and robustness of the search. The experimental diffraction pattern was optionally downsampled prior to peak detection and refinement to reduce computational cost while preserving the major diffraction features. An early stopping criterion was introduced to terminate the tree search once the refinement satisfied predefined quality criteria, thereby avoiding unnecessary exploration of additional branches. The phase scoring function was modified by normalizing the matched peak intensity to the total calculated intensity of the candidate phase rather than the total intensity of the experimental pattern, preventing phases with relatively few reflections from being systematically disadvantaged. The peak-matching algorithm was also modified to explicitly account for the widespread occurrence of overlapping reflections in powder X-ray diffraction. In the original implementation, calculated reflections that could not be matched to the residual peaks were classified as extra peaks and penalized. In the modified algorithm, such reflections are not considered extra if they coincide with reflections already explained by previously refined phases, thereby avoiding unnecessary penalties arising solely from peak overlap.

Candidate phases containing a larger number of diffraction peaks were preferentially refined before lower-information phases to improve search stability. Finally, when overfitting was detected after refinement of a parent node, expansion of all sibling branches at the same tree level was terminated to prevent cascading refinement of unlikely phase combinations. We also modified the grouping criterion: previously it was based on a chemistry-aware heuristic; it is now based purely on profile similarity (Gaussian-cosine clustering of synthetic peak lists), the same grouping convention used to score results throughout this work.

**5. Evaluation Metrics**

Throughout this work, we report micro-averaged precision, recall, and F1 as the primary evaluation metrics. Scoring is performed at the grouped-phase level described in Section 6, such that structurally near-identical candidate phases are treated as equivalent before true positives, false positives, and false negatives are aggregated across all patterns. Micro-averaging therefore weights individual phase occurrences equally. In contrast, macro-averaging assigns equal weight to each phase group regardless of how frequently it appears in the test set, while support-weighted averaging gives little or no weight to sparsely represented or absent phase groups. Micro-averaging is therefore better suited to our highly sparse multilabel setting, where each pattern contains only a small number of true phases relative to the full candidate library. The relevant equations are:

$$\mathrm{Precision_{micro}} = \frac{\sum_{i=1}^{N} \mathrm{TP}_i}{\sum_{i=1}^{N} (\mathrm{TP}_i + \mathrm{FP}_i)}$$

$$\mathrm{Recall_{micro}} = \frac{\sum_{i=1}^{N} \mathrm{TP}_i}{\sum_{i=1}^{N} (\mathrm{TP}_i + \mathrm{FN}_i)}$$

$$\mathrm{F1_{micro}} = \frac{\sum_{i=1}^{N} \mathrm{TP}_i}{\sum_{i=1}^{N} \mathrm{TP}_i + \frac{1}{2} \cdot \sum_{i=1}^{N} (\mathrm{FP}_i + \mathrm{FN}_i)}$$

where $\mathrm{TP}_i$, $\mathrm{FP}_i$, $\mathrm{FN}_i$ stand for true positive, false positive and false negative for the $i$-th pattern with $N$ patterns in total.

Top-$k$ accuracy on the pristine samples additionally credits a model if every true phase's group appears anywhere within its own top-k ranked candidates, which lets models with different native output formats (per-phase probabilities versus a single ranked list) be compared without adopting a single, non-comparable score threshold.

**6. Performance assessments based on grouped phases**

Scoring throughout this work uses the same profile-similarity grouping described earlier: the output of a list of phases is clustered at a 0.90 profile similarity threshold before comparison against the ground truth, since the 365-phase reference catalog contains many structurally near-identical phases that XRD alone struggles to distinguish. At this threshold, the 365 candidate CIFs cluster into 317 distinct groups (average size 1.15 CIFs/group), of which 32 contain more than one member (Supplementary Figure S3). Most of these multi-member groups span structurally near-identical phases of differing chemistry that a pure profile-similarity clustering correctly identifies as XRD-indistinguishable. For example, one 6-member group spans several Fe-O and Ti-Fe-O oxides, including $Fe_2O_3$ (space group 92), $Fe_{21.16}O_{31.92}$ (s.g. 212), $Fe_{21.332}O_{32}$ (s.g. 96), $Fe_{21.333}O_{32}$ (s.g. 213), $Fe_{62.92}O_{94.8}$ (s.g. 96), and $Ti_{3.36}Fe_{17.44}O_{32}$ (s.g. 212) (Supplementary Figure S4). Scoring at this group level, rather than by exact CIF identity, ensures a model is not penalized for naming a different, XRD-indistinguishable member of the same group as the true phase.

**7. Baseline models**

All four baseline classifiers were retrained from scratch on the same 365-phase reference catalog used throughout this work, since none ships with public pretrained weights for this chemical space. XQueryer[10] (github.com/Bin-Cao/XQueryer) architecturally expects an elemental-composition input *via* cross-attention; because no leakage-free composition is available at real inference time, every pattern is instead given the same fixed, pattern-independent composition prior (a mean-pooled CGCNN embedding over the catalog's constituent elements).

XCA[9] (github.com/maffettone/xca) was likewise retrained, patched only for TF-2.21/Keras-3 API compatibility. XRD-AutoAnalyzer[7,8] (github.com/njszym/XRD-AutoAnalyzer) was retrained against the same catalog. Because its native pipeline performs iterative greedy phase identification *via* full BGMN Rietveld refinement, which would duplicate DARA's own refinement step and contend for the same BGMN processes, the benchmark instead calls only its first CNN forward pass, which keeps the comparison to raw CNN baseline.

Peak Search-Match is performed using a custom implementation provided in our public GitHub repository. Peaks are detected *via* scipy's find_peaks function (prominence ≥5% of the maximum intensity, height ≥2%, minimum spacing 5 points) and converted to *d*-spacing *via* Bragg's law; reference patterns are pymatgen-simulated from the same 365 structure files and pre-cached; candidates are scored with a Smith/Snyder-style figure of merit, sweeping a ±0.1° zero-shift correction in 0.01° steps and requiring at least 6 total matches.

## 8. Sample Preparation

### 8.1 Chemicals

Chemicals used to construct the experimental test set include: Iron (II,III) oxide (95%, Sigma-Aldrich), iron (III) oxide (>96%, Sigma-Aldrich), lithium carbonate (>99%, Sigma-Aldrich), lithium iron phosphate (98%, aablocks), lithium manganese (III,IV) oxide (battery grade, Sigma-Aldrich), lithium phosphate (Sigma-Aldrich), lithium titanate (Sigma-Aldrich), manganese (II) carbonate (>99.9%, Sigma-Aldrich), manganese (II) oxide (99%, Sigma-Aldrich), manganese (II,III) oxide (97%, Sigma-Aldrich), manganese (III) oxide (99%, Sigma-Aldrich), manganese (IV) oxide (>99%, Sigma-Aldrich), titanium (IV) oxide, anatase (99.8%, Sigma-Aldrich), titanium (IV) oxide, rutile (99.9%, Sigma-Aldrich). The chemical for texture was Molybdenum(VI) oxide (>99.5%, Sigma-Aldrich).

### 8.2 XRD measurements

Most XRD measurements were done on a Shimadzu Lab X XRD-6000 with ANALIX A-40-Cu source. Powders were added directly onto a flat sample holder and leveled with a piece of glass, then measured from 10–80°. Measurement was done at a fast scan speed of 12°/min and step of 0.02° to produce data with a high degree of noise, thus representing a challenging test case for the methods evaluated here. Voltage and current were set to 40 kV and 35 mA respectively. A 1° divergence slit, 1° scattering slit and 0.15 mm receiving slit were used during the measurement.

For the tests involving sample displacement, measurements were done on a Bruker D8 Advance diffractometer with $K_{\alpha 1}$ = 1.54060 Å and $K_{\alpha 2}$ = 1.54439 Å. Samples were placed on an alumina holder and measured from 10–80° with 0.020° step size and 0.10 s/step scanning rate. Voltage and current were set to 40.0 kV and 40.0 mA, respectively. A divergence slit of 0.6 mm was used.

### 8.3 Preparation of mixtures and artifact-containing samples

Multiphase mixtures were prepared using a Hauschild SpeedMixer DAC 150.3 FVZ with zirconia balls, then loaded onto the same flat sample holder for Shimadzu measurement. For the impurity-containing samples, powders were weighed to reach minority-phase weight fractions of 1, 2, 3, 5, 10, 15, and 20 wt%, then mixed *via* SpeedMixer and measured on the Shimadzu XRD.

For uniformly peak-shifted patterns, previously measured single-phase patterns were shifted by 0, 0.1, 0.2, 0.3, 0.4, 0.5, 0.75, and 1.0°, with the resulting missing regions at the pattern edges filled by interpolation. To probe sample displacement, single-phase samples were measured on a Bruker D8 Advance diffractometer, with sample displacement mimicked by offsetting the sample height $z$ by 0, 0.1, 0.2, 0.3, 0.5, 0.75, 1.0, and 1.5 mm from the focusing circle. To obtain peak broadening, samples were milled in a Retsch PM 100 planetary ball mill: 10 g of sample and 30 steel balls were combined with an appropriate amount of ethanol and milled for a total of 1 h at 500 rpm, alternating 10 min of milling with 10 min of rest, prior to XRD measurements.

To probe texture, 2D $MoO_3$ nanosheets were prepared following a prior report:[37] 200 mg of $MoO_3$ was dispersed in 5 mL of ethanol and sonicated for 10 min. To obtain a fully textured pattern, 200 µL of the resulting supernatant was drop-cast directly onto a glass slide on a 70 °C hotplate. To reduce texture, the supernatant was first milled using a SpeedMixer at 3500 rpm before an identical drop-cast step. Varying this pre-drop-cast milling time produced samples with a range of texture magnitudes, with longer milling times producing progressively less texture.

### 8.4 ***In-situ*** XRD data collection

*In-situ* data were collected on a Bruker D8 Advance diffractometer (Cu K$\alpha_1$, $\lambda$ = 1.5406 Å) fitted with an Anton Paar HTK 1200N hot stage. Each reaction began at 50 °C and was heated in 25 °C increments at 5 °C/min. At each temperature increment, heating was paused while a diffraction pattern was collected over $2\theta$ = 10–80° (~20 min per scan). This sequence was repeated until the final temperature was reached: 800 °C for $BiFeO_3$, 950 °C for $CaTiO_3$, and 1100 °C for $CaCu_3Ti_4O_{12}$. The resulting datasets contained 31, 37, and 43 patterns, respectively.

# Supplementary Information

# Scalable machine learning framework for multiphase identification from powder X-ray diffraction

Xinyang Tong,[1] Ethan Jin,[2] Jiahan Xu,[1] Aditya Rao,[3] Pengcen Jiang,[4]
and Nathan J. Szymanski[1,*]

**Table of Contents**

[1] University of California, Los Angeles, Department of Materials Science & Engineering, Los Angeles, CA 90095
[2] University of California, Los Angeles, Electrical and Computer Engineering Department, Los Angeles, CA 90095
[3] University of California, Los Angeles, Mechanical and Aerospace Engineering Department, Los Angeles, CA 90095
[4] University of California, San Diego, Halıcıoğlu Data Science Institute, La Jolla, CA 92093
[*] Correspondence to nszymanski@seas.ucla.edu

## Supplementary Note S1. The role of downstream refinement

GALAXI's phase-specific CNN classifiers are trained independently of one another. Each network is a binary detector for a single target phase, with no awareness of which other phases might also be present in the same pattern. This independence allows the library to scale without retraining the existing classifiers, but it also makes the CNN screen intentionally over-inclusive. Because diffraction intensities from multiple phases add together in a real mixture, an independent classifier cannot determine whether adding a particular phase to the other flagged phases improves the explanation of the observed pattern. It can only assess whether the pattern is consistent with that phase being present in isolation.

DARA supplies this missing combination-level reasoning. It evaluates candidate phase combinations against the full experimental diffraction profile through Rietveld refinement, allowing phases that expand the CNN-flagged candidate set without improving the overall fit to be discarded. Thus, the CNNs provide scalable phase-level screening, while DARA determines which combinations of the screened phases are consistent with the complete diffraction pattern.

An alternative would be to replace the independent classifiers with a single correlated multi-label model that reasons jointly over all phases. However, this would introduce dependencies among phase labels and undermine the decoupled design that allows GALAXI to add or update phase-specific classifiers independently. Such a model would need to learn phase-combination behavior directly, rather than making one independent decision per phase. DARA instead preserves the scalability of independent phase-specific screening while providing the combination-level validation required for real multiphase diffraction patterns.

**Supplementary Note S2. Synthetic patterns used for training**

Before any exposure to experimental data, each phase-specific network is trained and evaluated exclusively on synthetic patterns generated by the artifact simulating pipeline. We benchmark this on a 200-phase reference catalog sampled broadly from the COD to span a wide range of chemistries and structure types, rather than being restricted to a single narrow chemical space. Across these 200 phase-specific models, the ensemble reaches a mean held-out synthetic test accuracy of 98.0% ± 0.8% and a mean test area under the ROC curve (AUC) of 0.995 ± 0.003, with the single worst-performing model still exceeding 94.1% accuracy. This uniformly high, low-variance performance was maintained across the range of architectures explored during hyperparameter tuning, indicating that, given a sufficiently large and diverse synthetic training set, the one-versus-all binary classification task is comparatively easy for a CNN to learn across this broad and chemically diverse reference catalog. On simulated patterns alone, model capacity is not the limiting factor for this task.

**Supplementary Note S3. Experimental validation set for hyperparameter tuning**

The goal of this hyperparameter tuning was not to maximize F1 in isolation, but to reach a high-recall, acceptable-precision operating point. Because the downstream DARA refinement stage (Methods Section 4) is well-suited to filtering false positives out of an over-inclusive candidate list, we weigh recall more heavily than precision when selecting pattern-generation and model-configuration hyperparameters. A missed phase can never be recovered by refinement, while an extra candidate can often be discarded by it. Precision cannot be neglected entirely, however, since an excessively noisy candidate list gives DARA more opportunities to overfit onto a spurious lookalike phase. The hyperparameters reported in the main text were therefore chosen to keep recall high while keeping precision in a regime DARA can reliably clean up. Full results for every parameter swept, one factor at a time, against the 57-pattern experimental validation set are given below.

**Note S3.1 Experimental validation set source**

Patterns for the experimental validation set were collected from a combination of three prior works.[1–3] The set spans a wide list of chemical spaces: Bi-O, Co-O, Cr-O, F-Li, F-Mn, In-O, Mn-O, Ni-O, O-Ti, O-V, O-Zn, O-Zr, C-Co-O, C-Li-O, C-Mn-O, C-Na-O, H-La-O, H-Li-O, Li-Mn-O, Li-O-P, Li-O-Ti, N-Na-O, Al-Li-O-Si. The full dataset can be found in the reproducing/validation folder of https://github.com/Szymanski-Group/galaxi.

**Note S3.2 Pattern-generation parameter sweeps**

The uniform peak-shift range (±0.30°) sits at a clear F1 peak (micro-F1 = 0.770), degrading on either side (±0.05°: 0.731; ±1.00°: 0.607) — too narrow a range leaves the model unable to recognize genuinely present phases whose peaks are shifted beyond the training distribution, while too wide a range teaches the model to tolerate large angular shifts indiscriminately and inflates false positives.

Sample displacement showed a similar, if shallower, trade-off: the narrowest tested range (±0.10 mm) slightly underperformed (micro-F1 = 0.747), while the ±0.50 mm range gave the highest performance (micro-F1 = 0.770). Expanding the range further to ±1.00 and ±1.50 mm reduced micro-F1 to 0.737 in both cases.

Raising the crystallite-size lower bound (which sets how much peak broadening is simulated from small, nanocrystalline domains) steadily improved precision (0.60 to 0.71) and micro-F1 (0.744 to 0.805) as the bound rose from 0.1 to 50 nm, but at a real recall cost at the widest bound tested: recall drops to 0.928 at [50, 450] nm, versus 0.97-0.98 at any narrower bound. Since our tuning goal favors recall, this F1 gain is not a pure win — excluding small crystallite sizes too aggressively risks missing genuinely broadened, small-crystallite real phases.

Lattice strain showed a sharper, one-sided trend: narrowing the strain range to ±0.001 performed comparably to the frozen baseline (micro-F1 0.775), but widening it to ±0.01, ±0.02, or ±0.05 caused a steep, accelerating collapse (F1 0.740, 0.628, and 0.476, respectively), with precision falling from 0.65 to 0.33 as excessive simulated strain increasingly overlaps with genuine chemical and structural differences between phases.

Microstrain showed a similar one-sided sensitivity to lattice strain, though over a much smaller absolute range: narrowing the microstrain upper bound to 0.001 outperformed a wider 0.004 bound (micro-F1 0.791 vs. 0.734), with the narrower setting improving precision (0.67 vs. 0.59) at a comparable recall (0.97-0.98). Excessive simulated microstrain broadens and distorts peak shapes enough to blur the boundary between distinct phases, so keeping this parameter on the conservative side improves discrimination without a meaningful recall cost.

The target-phase weight-fraction lower bound traced the clearest precision-recall trade-off of any parameter, and its optimum is a genuine compromise rather than a value that is simply best left unperturbed. Raising the lower bound from 0.01 to 0.10 to 0.20 to 0.35 monotonically traded recall for precision (recall 0.99 → 0.97 → 0.94 → 0.84; precision 0.55 → 0.67 → 0.75 → 0.82), with micro-F1 rising from 0.707 to 0.791 to a peak of 0.832 at a 0.20 lower bound before edging back down to 0.830 at 0.35. Given our preference for recall over precision in this screening stage, the frozen configuration retains a low lower bound — below the F1-optimal 0.20 — deliberately trading some precision for the higher recall that keeps genuinely present, low-abundance phases from being missed before they ever reach DARA.

Noise level proved to be one of the most robust parameters tested: even at a simulated noise level as high as 10%, micro-F1 (0.771) and recall (0.973) remained close to their values at much lower noise (e.g. 0.756 F1 at 0.5% noise), indicating the model is largely insensitive to noise magnitude over this range. Precision and recall did, however, trace opposite trends with noise level for a clear mechanistic reason. At low noise, the model can learn to rely on exact peak positions and unambiguously distinguish even small position differences between phases, giving high precision; but because real experimental noise can then push a genuine peak below the model's learned confidence threshold, recall suffers. At high noise, the model instead learns during training to become conservative and attend only to the most prominent peaks, which raises precision by avoiding noise-driven false positives; however, if a mixture happens to suppress those dominant peaks (for example, a low-abundance phase whose signature reflections are otherwise strong), the model becomes more likely to miss it, lowering recall. The two effects roughly balance across the tested range, which is why overall F1 stays comparatively flat.

Texture magnitude, by contrast, showed no beneficial range at all: disabling simulated texture entirely gave the best validation performance (micro-F1 0.777), and every wider texture range we tested made performance monotonically worse (F1 0.746 at [0.75, 1.25], 0.724 at [0.25, 2], and 0.499 at the widest range tested, [0.1, 10]). We nonetheless retain a moderate texture range ([0.5, 1.5]) in the frozen configuration rather than disabling texture outright, since texture is not uncommon in real bulk powder samples, and the validation-set cost of a moderate range is small relative to the risk of a model with no texture tolerance at all.

**Note S3.3 Preprocessing**

On the preprocessing side, we confirmed that the three-stage pipeline (smoothing, background subtraction, and noise-floor gating with non-linear peak magnification) is essential, and each of its hyperparameters has a distinct, mechanistically interpretable effect. The first stage, Savitzky-Golay smoothing, is controlled by its window length: our sweeps show a clear optimum near the frozen 21-point window (micro-F1 = 0.770), degrading on both sides but more steeply toward wider windows (F1 = 0.754 at 11 points, 0.737 at 5 points, versus a sharper drop to 0.701 at 31 points and 0.661 at 41 points). Too narrow a window leaves residual noise for later stages to handle, while too wide a window over-smooths genuine peak shape and position, blurring closely spaced peaks together and directly hurting precision.

Noise sensitivity, applied after background subtraction, scales the signal's estimated noise level to set the baseline intensity above which a data point is trusted as real signal rather than noise. Gate sharpness then controls how hard the cutoff is at that boundary. A small gate sharpness makes the noise/signal cutoff smoother, so more noise "escapes" the threshold and gets amplified along with genuine peaks during training, and the model responds by learning to avoid noisy regions and rely only on strong, unambiguous peaks – favoring high precision at some cost to recall. A large gate sharpness instead removes nearly all noise before training, so the model learns from very clean patterns and becomes more easily fooled by overlapping peaks or residual noise at inference on real, messier data.

Finally, the non-linear power-law magnification step exists specifically to amplify weak peaks relative to strong ones before the network ever sees the pattern. Its exponent was the single most sensitive preprocessing hyperparameter in our sweeps, tracing a sharp inverted-U around its frozen value – too little magnification leaves minor phases buried under stronger reflections, while too much magnification over-amplifies residual noise into apparent peaks. Under- or over-tuning any of these stages sharply degraded validation performance even when simulated-pattern accuracy remained high, because real XRD intensities span a much wider dynamic range than idealized simulated patterns and weak, diagnostic peaks are otherwise lost in the background.

**Note S3.4 Model configuration**

The full hyperparameter grid for convolutional channel width and fully connected layer size showed that a [32,32,32] convolutional configuration with a 32-unit fully connected layer achieved the highest micro-F1 (0.8105). However, because our approach requires maintaining one independently trained model per candidate phase, we retained the smaller [8,8,8] convolutional configuration with an 8-unit fully connected layer for the final model, favoring reduced training time and storage cost.

**Note S3.5 Mask-modulated attention benefit**

We evaluated the benefit of mask-modulated attention using 57 experimental XRD patterns and 200 reference phases, with DARA disabled. The no-mask baseline was trained on the standard 10–80° range, whereas the mask-modulated model used throughout this work was trained on a broader 5–105° range with random masking. Both models were evaluated on the same 10–80° patterns using micro-averaged precision, recall, and F1.

The model with mask-modulated attention substantially outperformed the no-mask baseline, increasing micro-F1 from 0.680 to 0.770, with improvements in both precision (0.530 → 0.634) and recall (0.946 → 0.982). This demonstrates that mask-modulated attention helps the CNN focus on informative diffraction peaks rather than treating the entire angular range equally, motivating its use throughout this work.

## Supplementary Note S4. Additional results from *in-situ* measurements

### Note S4.1 $CaCO_3 + TiO_2 \rightarrow CaTiO_3 + CO_2$

We start with a simpler case as shown in Supplementary Figure S6: a $CaCO_3$ + $TiO_2$ → $CaTiO_3$ + $CO_2$ solid-state reaction (37 patterns, ramped from 50 to 950 °C). The $CaCO_3$ and $TiO_2$ precursor phases persist largely unreacted through 700 °C; our framework then identifies the target $CaTiO_3$ perovskite emerging at 725 °C and rising steadily to a weight fraction of 40.8% by 950 °C (still increasing at the ramp ceiling, consistent with an incomplete reaction). Decomposition of $CaCO_3$ begins around 800 °C and completes by 850 °C, with the resulting CaO transiently peaking at 35.2% weight fraction at 850 °C before declining to 25.1% by 950 °C as it continues reacting with residual $TiO_2$ to form additional $CaTiO_3$.

### Note S4.2 $CaCO_3 + 3\ CuO + 4\ TiO_2 \rightarrow CaCu_3Ti_4O_{12} + CO_2$

A more complex example is the $CaCO_3$ + 3 CuO + 4 $TiO_2$ → $CaCu_3Ti_4O_{12}$ + $CO_2$ reaction (43 patterns, 50 to 1100 °C) shown in Supplementary Figure S7. We first observe decarbonation of $CaCO_3$ to CaO beginning at 725 °C; the target $CaCu_3Ti_4O_{12}$ phase first appears at 750 °C and grows steadily to a phase-pure 100% weight fraction by 1025 °C. Along this pathway, our framework also correctly identifies small amounts of the intermediates $CaTiO_3$ (4.4% weight fraction, 975 °C) and $Ca_4Ti_3O_{10}$ (4.1%, 1000 °C), consistent with a reaction pathway proceeding through transient CaO + $TiO_2$ intermediates before full conversion to the target phase.

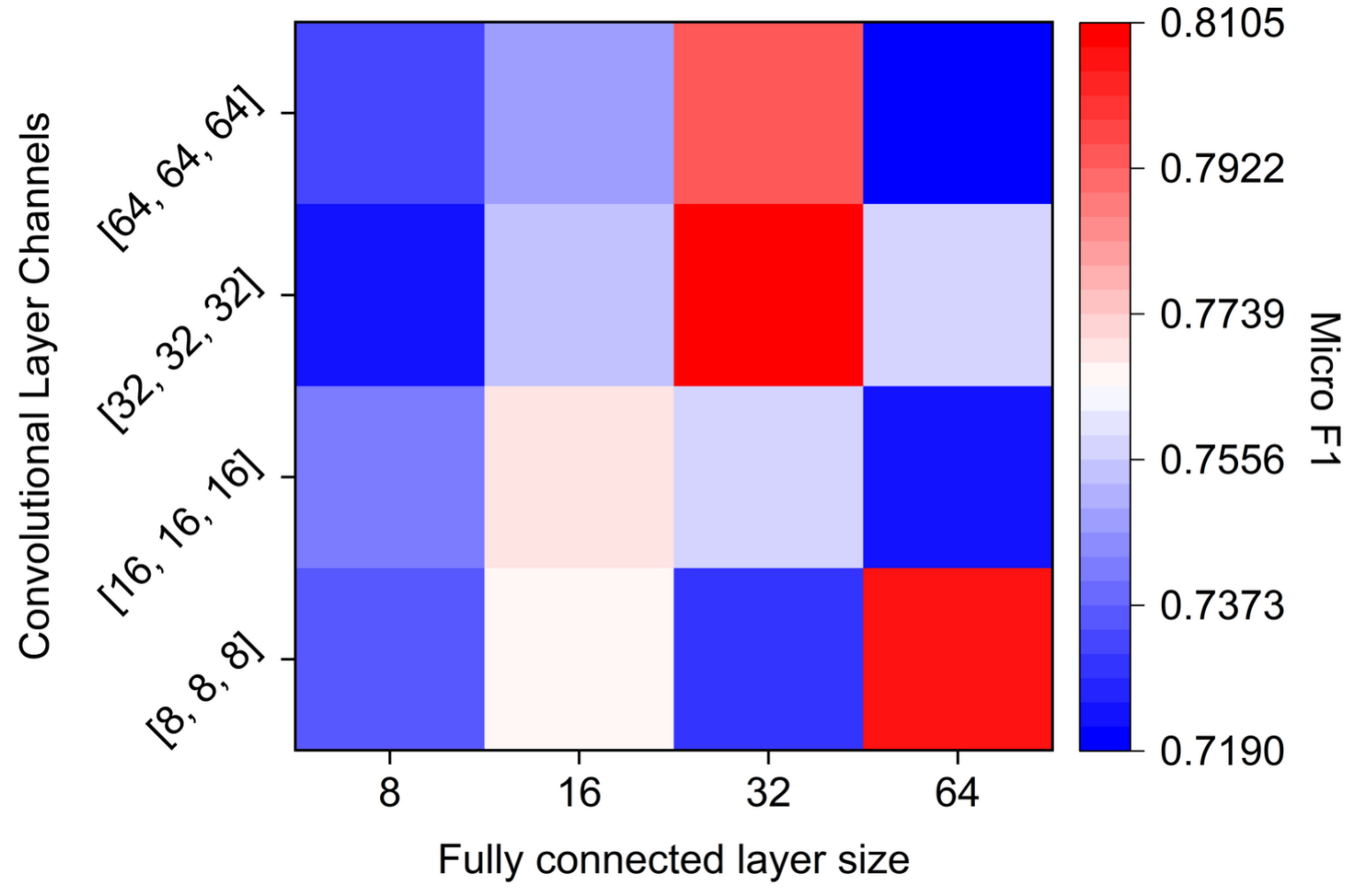

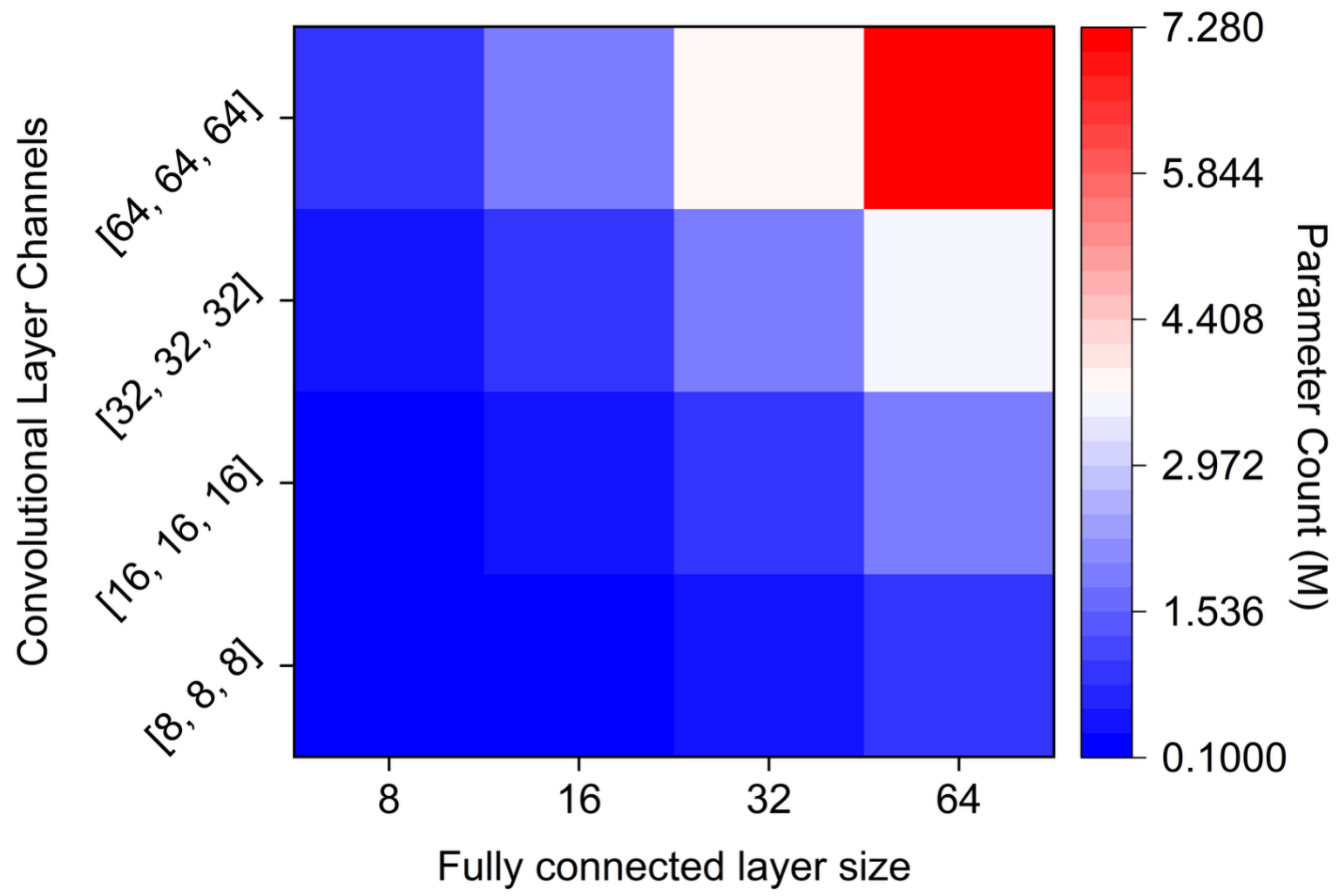

**Supplementary Figure S1. Hyperparameter grid evaluated against the 57-pattern experimental validation set.** Top: Micro-F1 score as a function of convolutional-channel width (rows) and fully-connected-layer size (columns). Bottom: Corresponding parameter count (in millions) for each configuration.

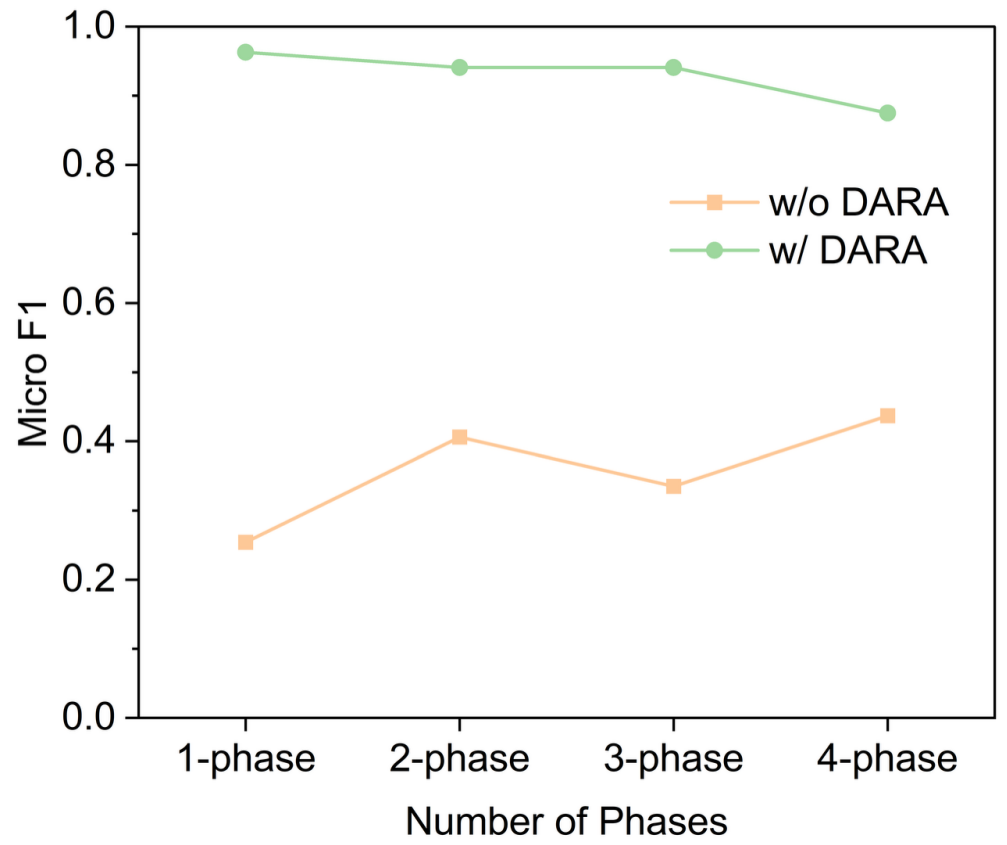

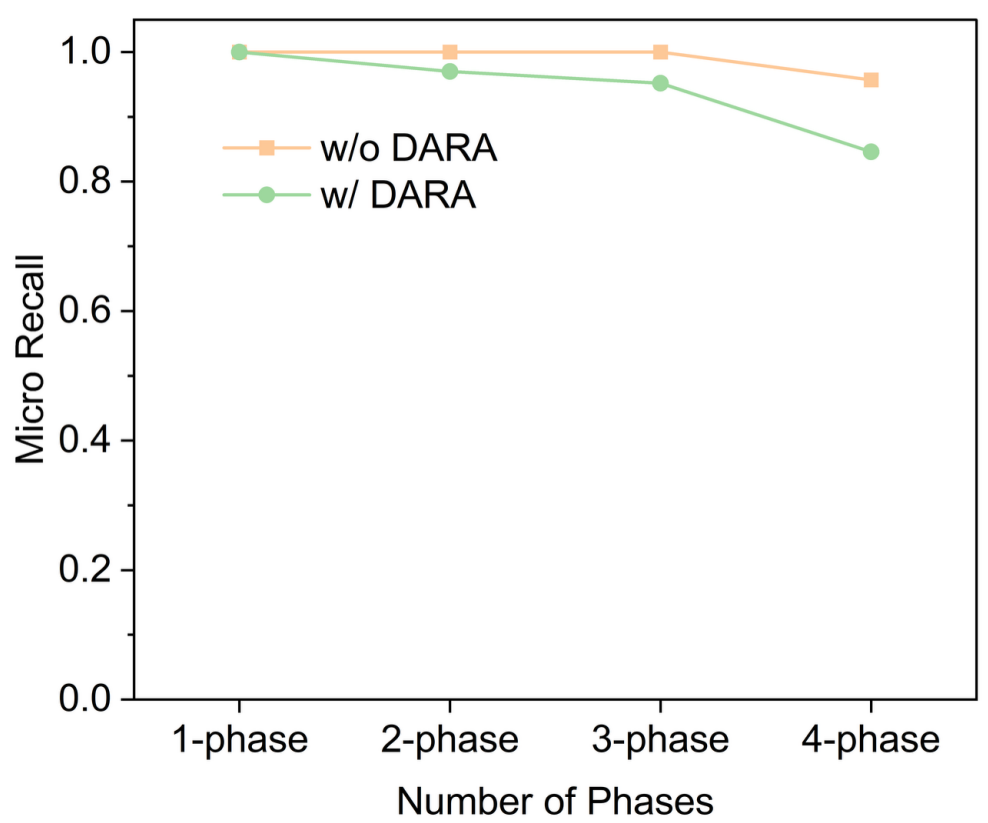

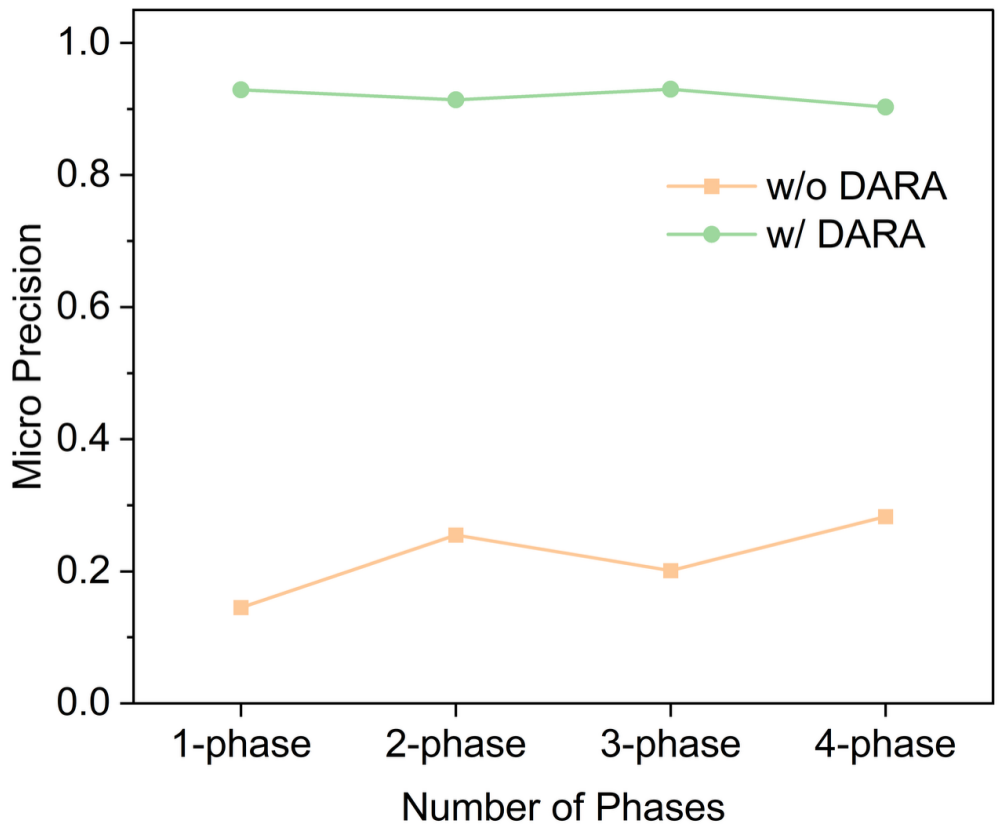

**Supplementary Figure S2.** GALAXI's micro-F1 score, recall, and precision on the pristine set of experimental patterns, with and without DARA, as a function of the number of true phases in each sample.

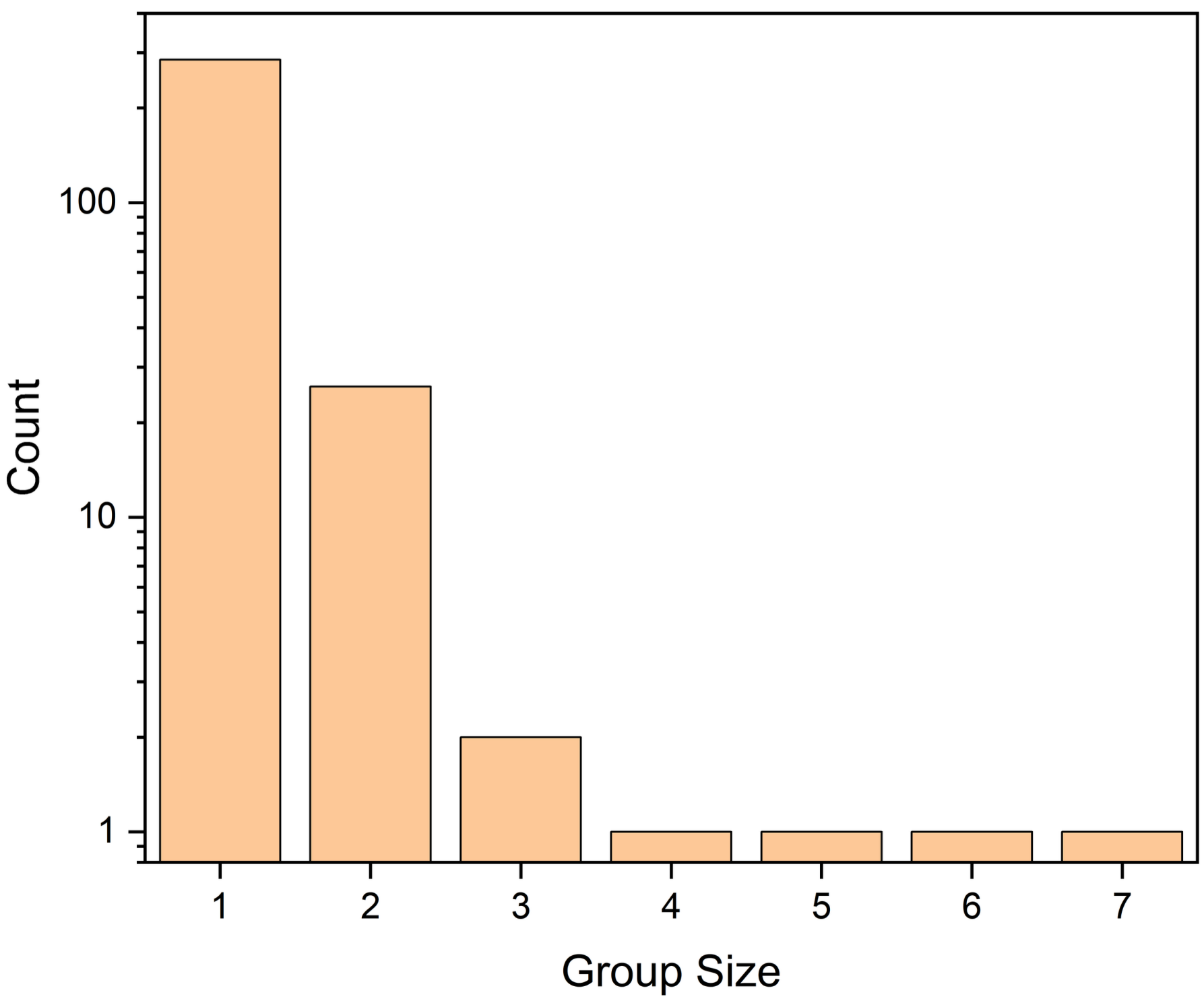

**Supplementary Figure S3.** Distribution of group sizes when the 365-phase reference catalog is clustered at the 0.90 profile-similarity threshold used throughout this work.

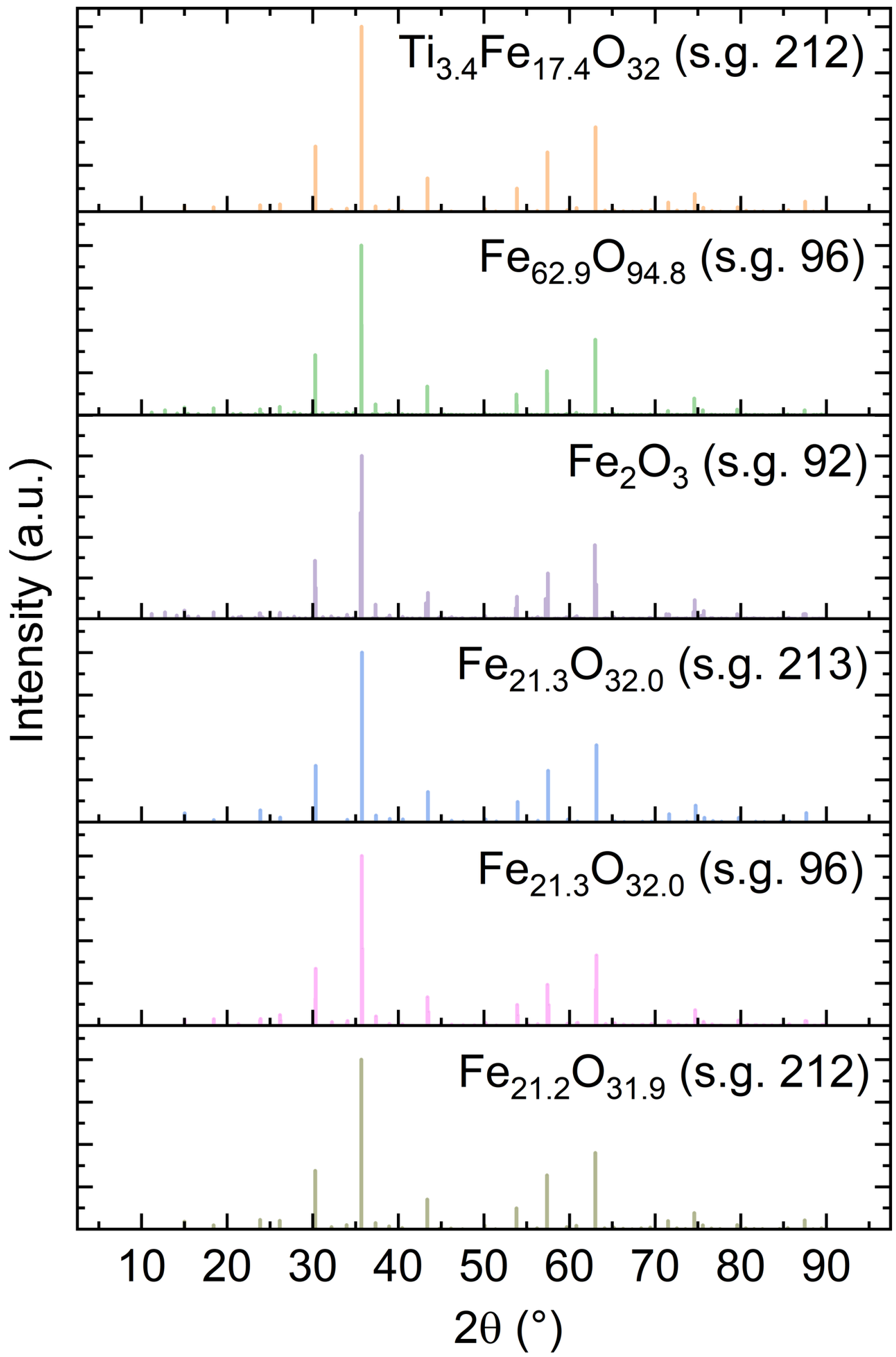

**Supplementary Figure S4.** Simulated stick patterns for the six members of one representative multi-phase similarity group (Fe-O and Ti-Fe-O oxide polymorphs), illustrating why they are XRD-indistinguishable at the 0.90 grouping threshold.

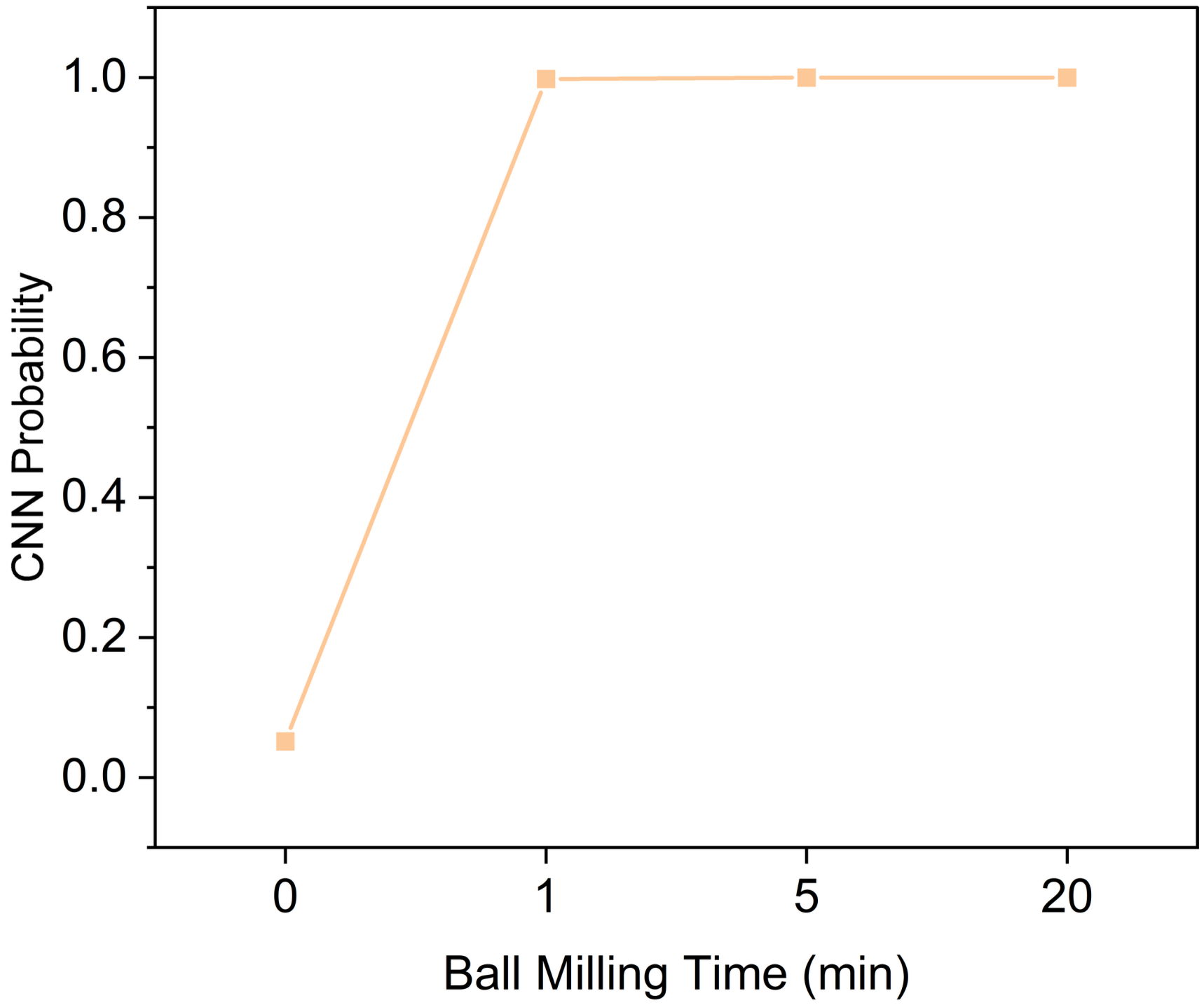

**Supplementary Figure S5.** GALAXI's raw CNN probability for $MoO_3$ as a function of pre-drop-cast SpeedMixer milling time, illustrating the texture diagnostic.

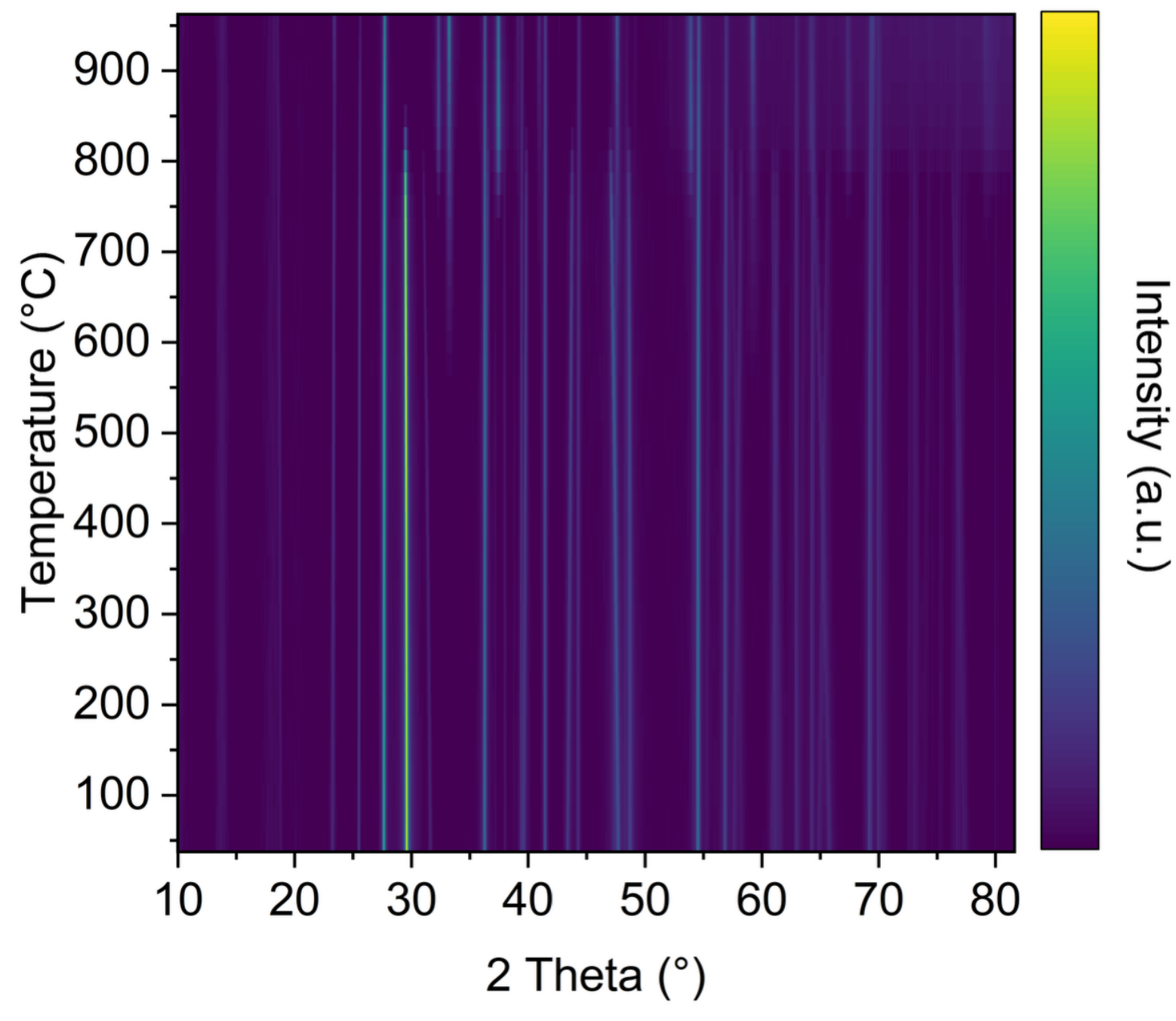

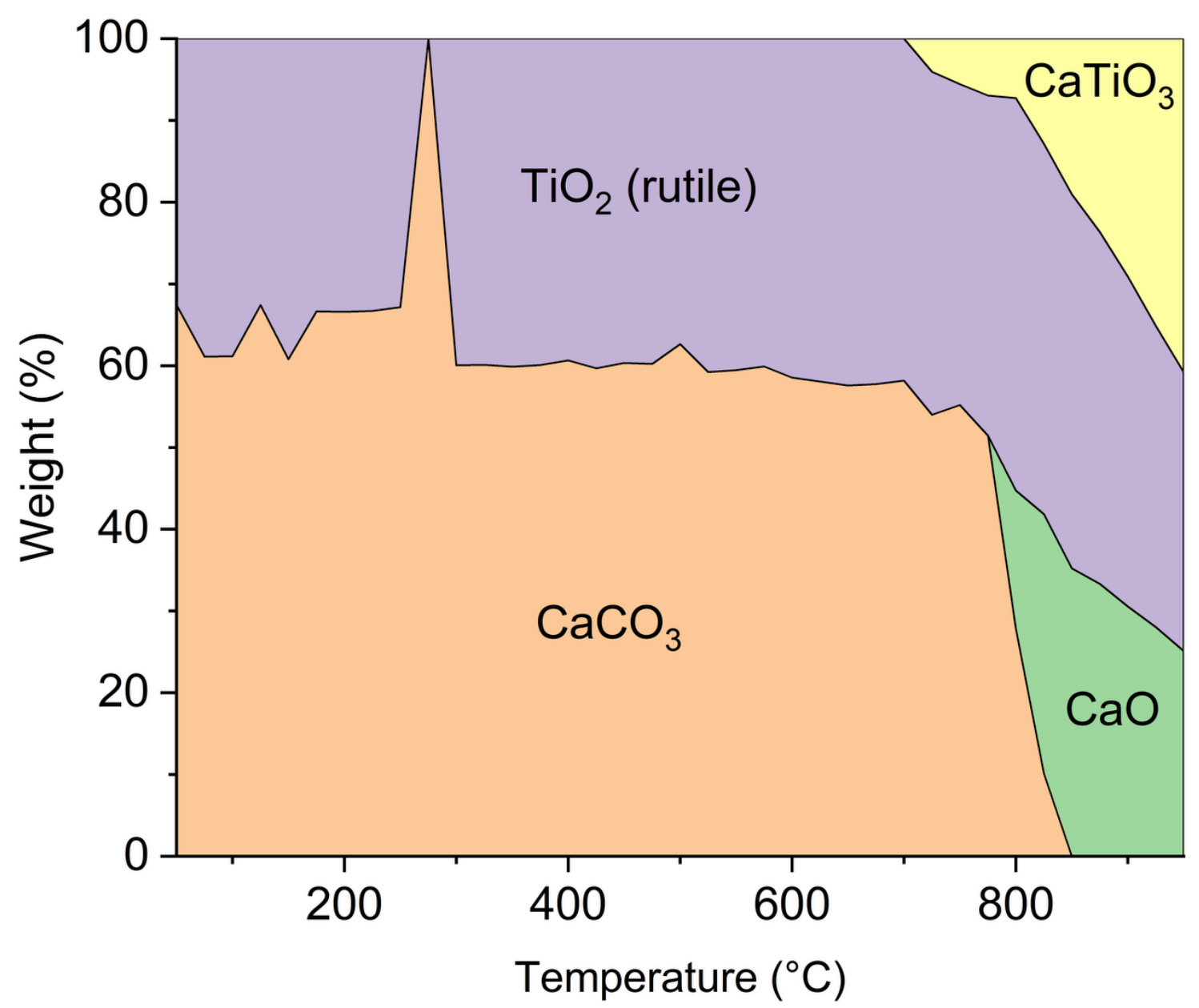

**Supplementary Figure S6.** Time-resolved *in-situ* XRD for the $CaCO_3 + TiO_2 \rightarrow CaTiO_3 + CO_2$ reaction. Top: Raw diffraction-intensity heatmap as a function of temperature and 2θ. Bottom: GALAXI-predicted phases and weight fractions versus temperature.

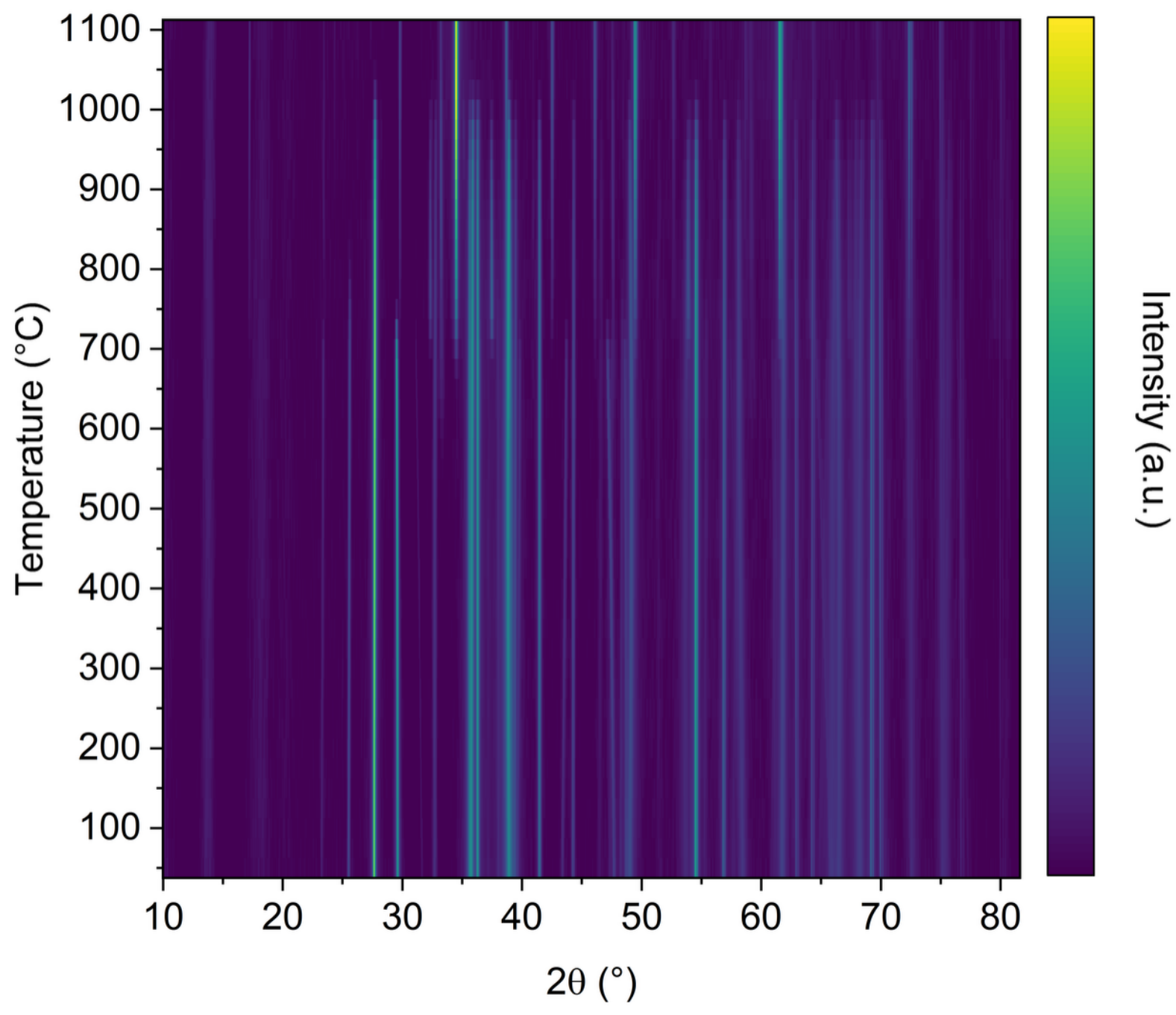

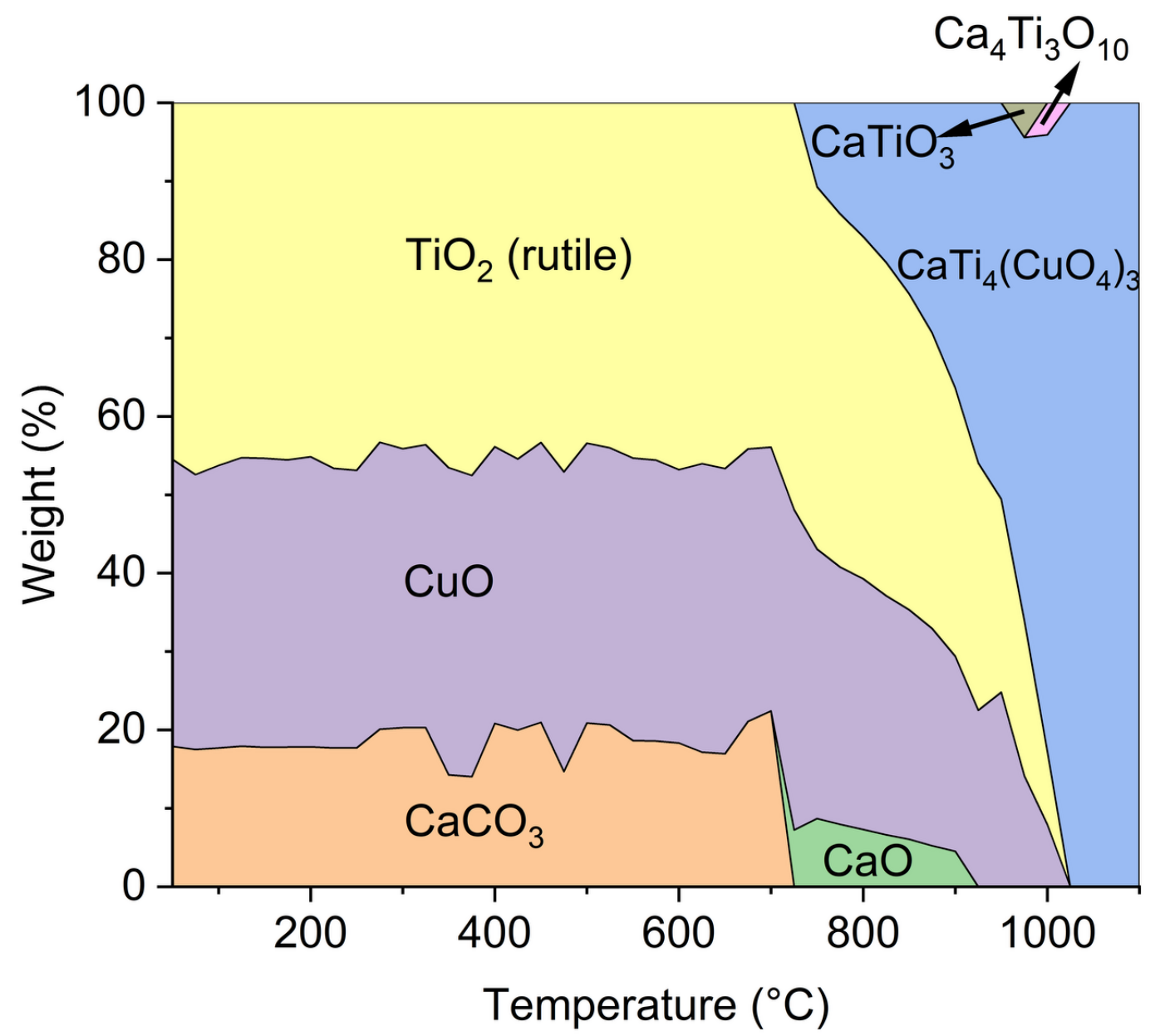

**Supplementary Figure S7.** Time-resolved *in-situ* XRD for the $CaCO_3$ + 3 CuO + 4 $TiO_2$ → $CaCu_3Ti_4O_{12}$ + $CO_2$ reaction. Top: Diffraction-intensity heatmap as a function of temperature and 2θ. Bottom: GALAXI-predicted phases and weight fractions versus temperature.

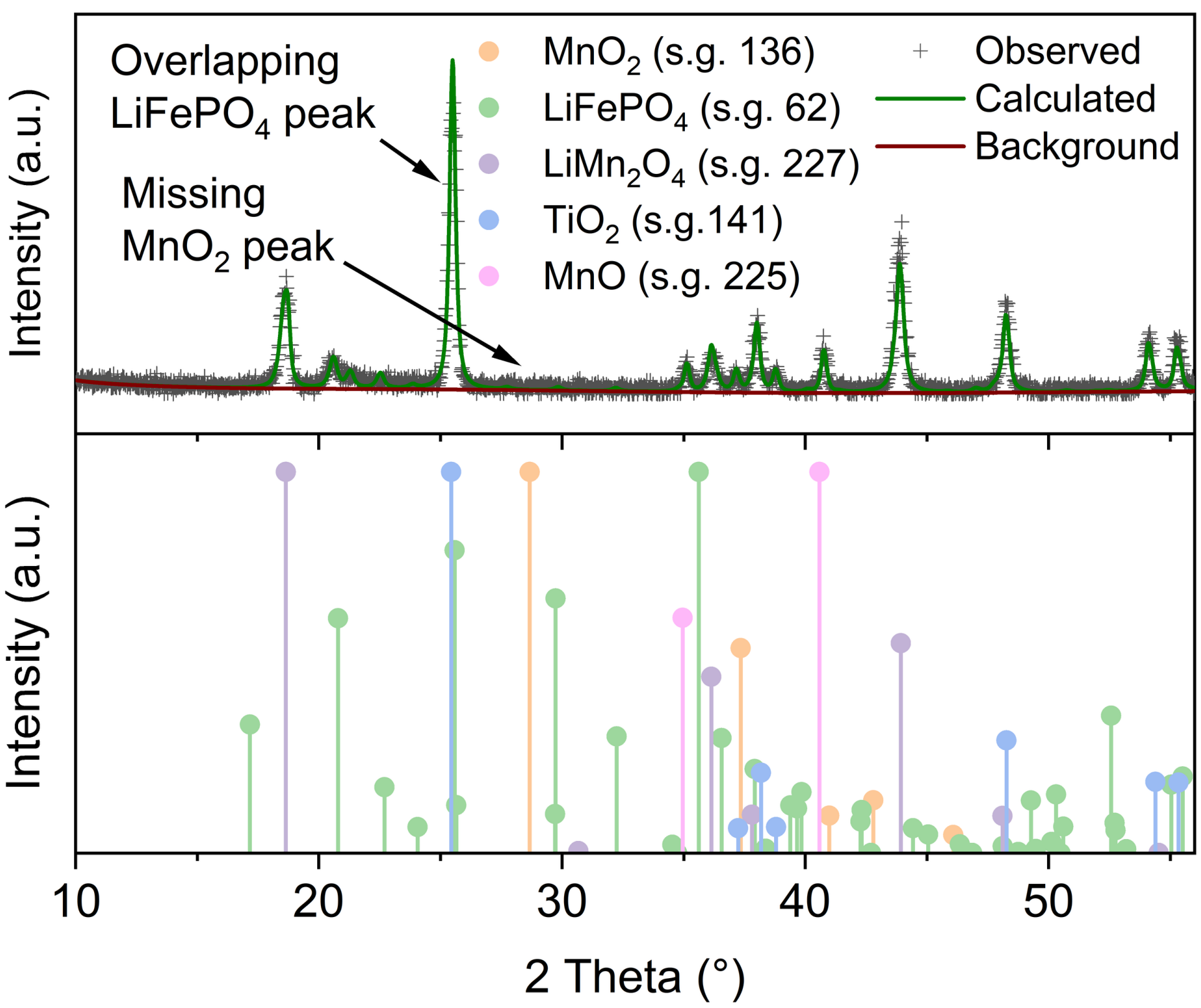

**Supplementary Figure S8.** DARA's refined fit for the four-phase $MnO_2$ + $LiFePO_4$ + $LiMn_2O_4$ + $TiO_2$ mixture, annotated with the missing $MnO_2$ peak and the $LiFePO_4$ peak overlapping it.

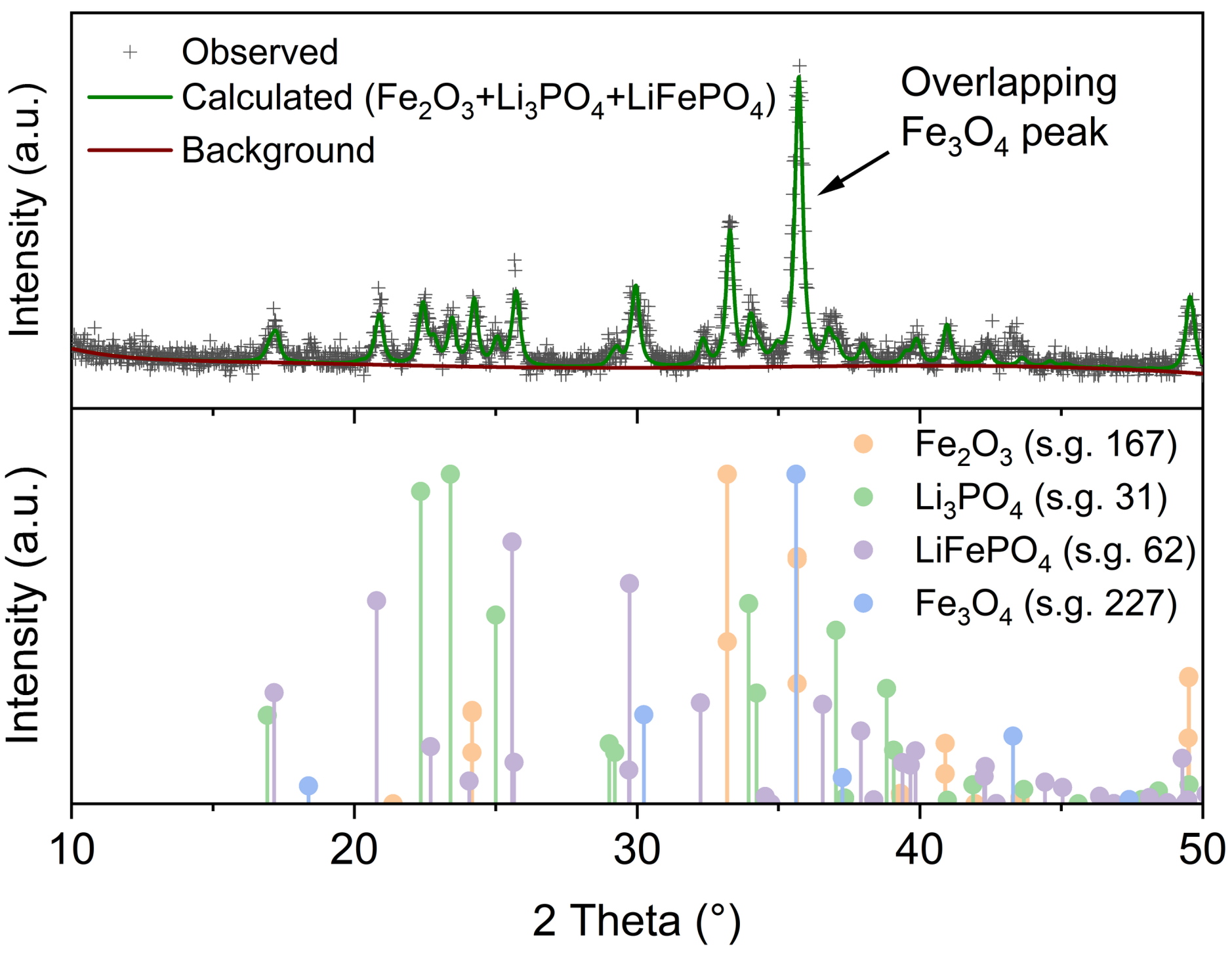

**Supplementary Figure S9.** DARA's refined fit for the four-phase $Fe_2O_3$ + $Fe_3O_4$ + $Li_3PO_4$ + $LiFePO_4$ mixture, annotated with the $Fe_3O_4$ peak overlapping the $Fe_2O_3$ peak.

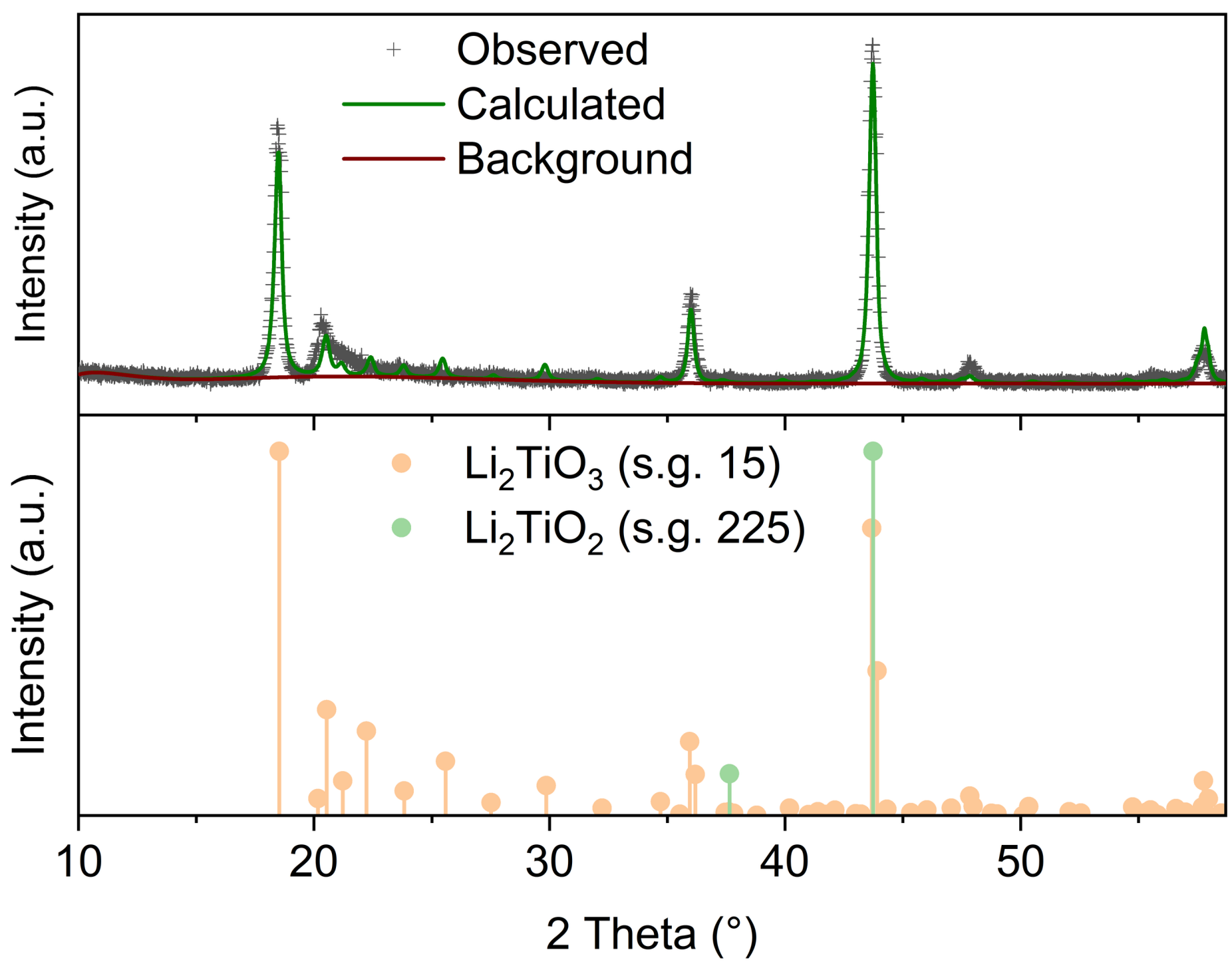

**Supplementary Figure S10.** DARA's refined fit for the single-phase $Li_2TiO_3$ pattern, showing the structurally similar $LiTiO_2$ phase DARA retains alongside the true $Li_2TiO_3$ phase.

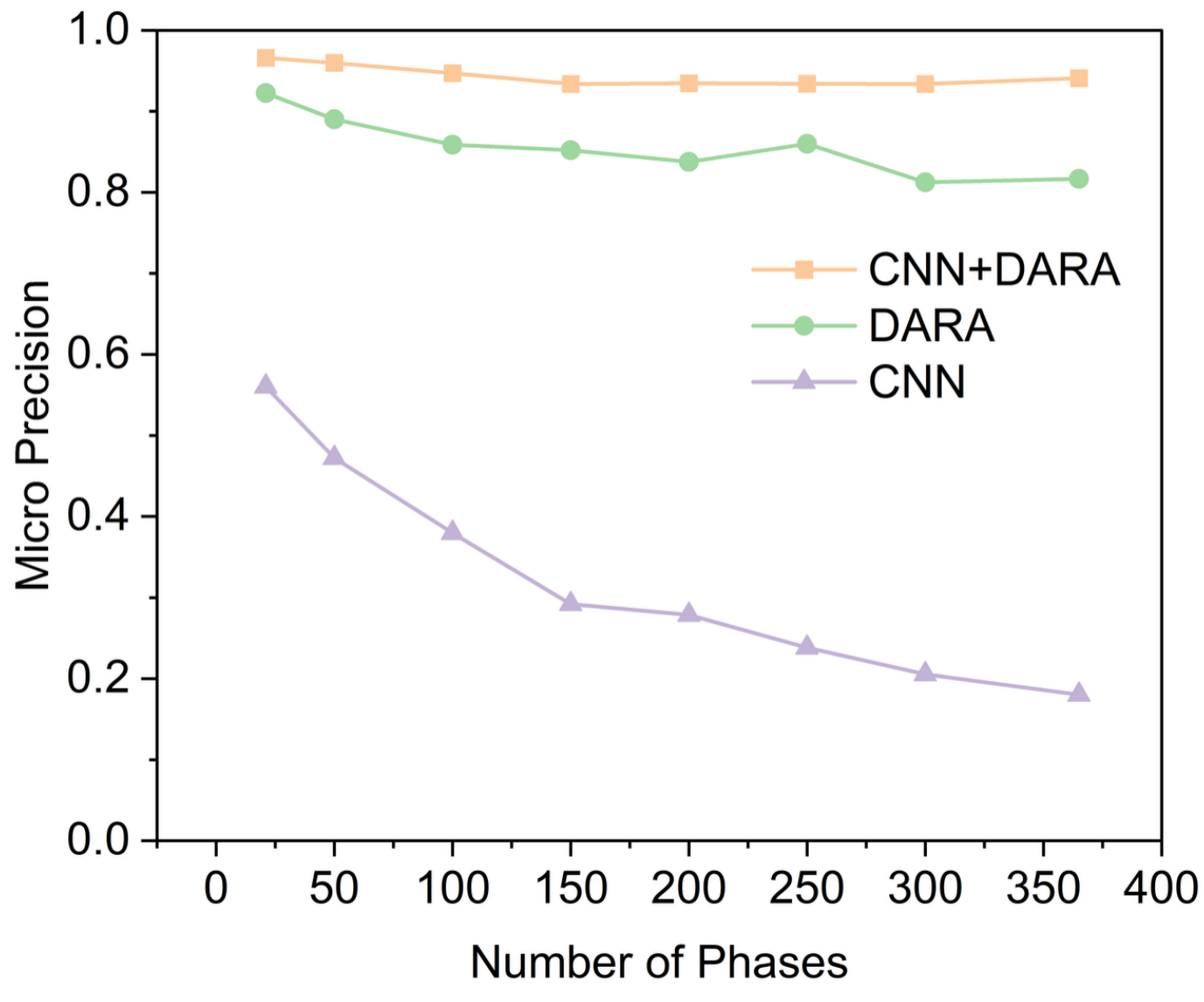

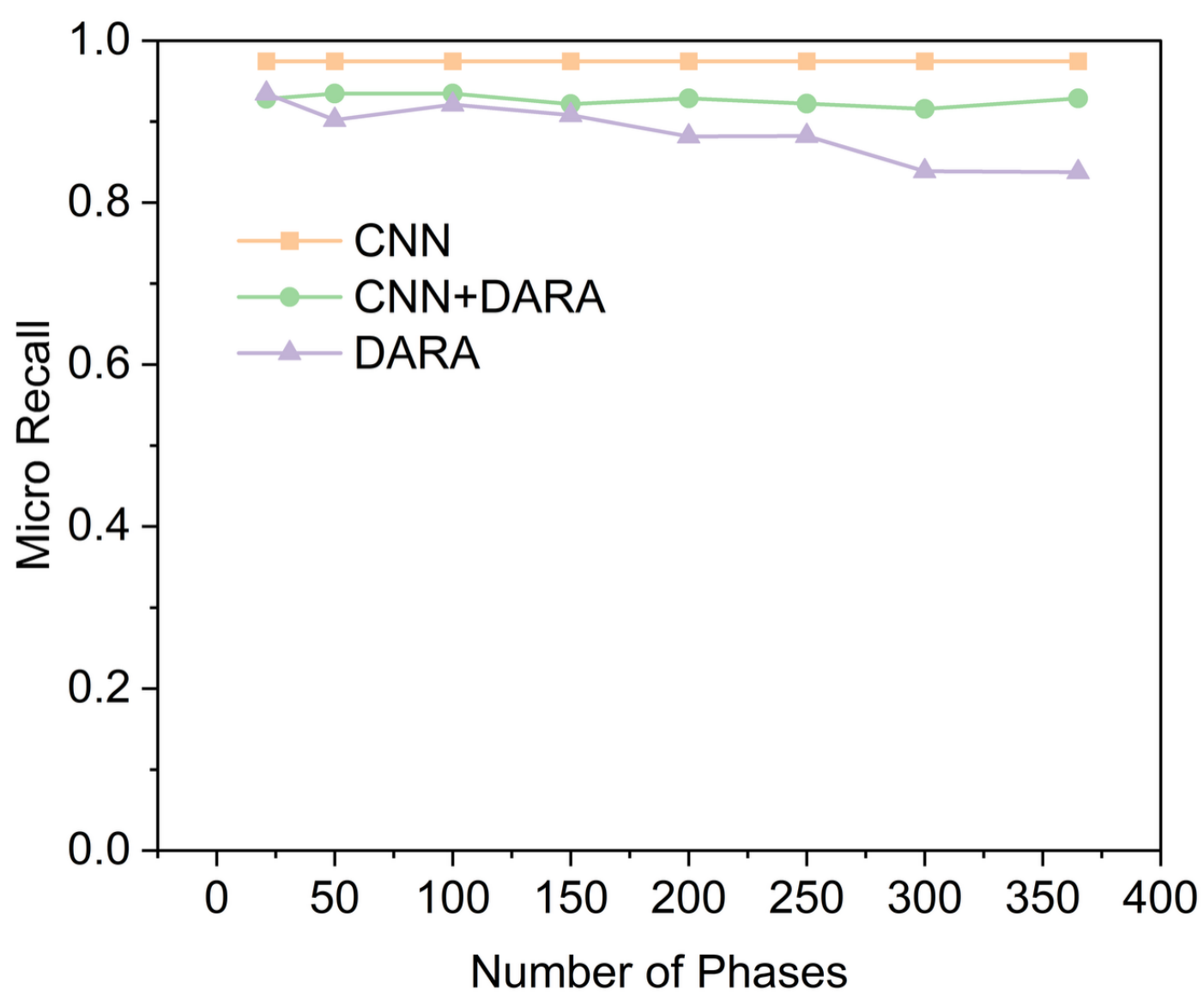

**Supplementary Figure S11.** Scaling of GALAXI's micro precision and recall with reference-catalog size on the 61-pattern pristine test set. Top: Micro-Precision plotted against the number of candidate phases for CNN-only, CNN+DARA and DARA-only models. Bottom: Micro-Recall plotted against the number of candidate phases for CNN-only, CNN+DARA and DARA-only models.

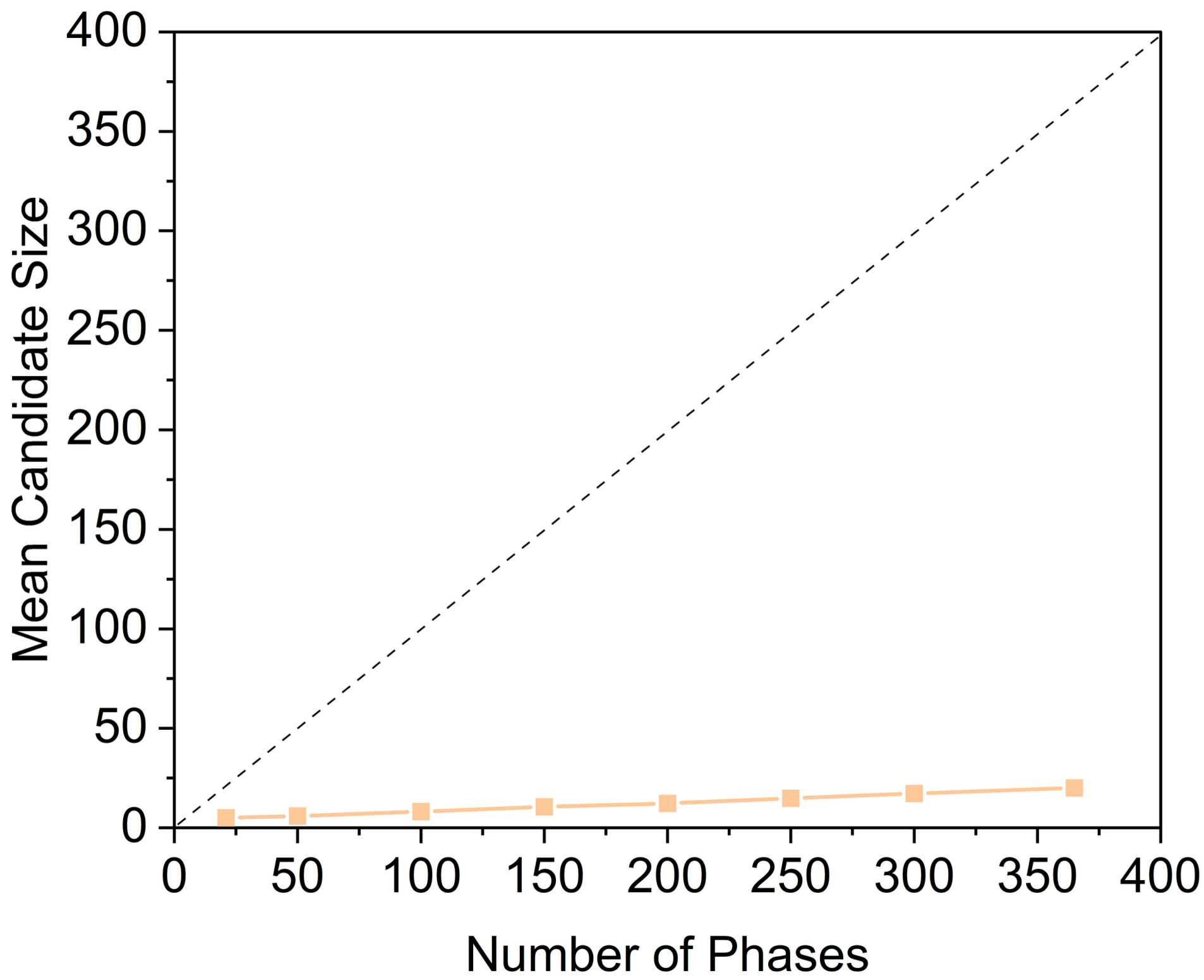

**Supplementary Figure S12.** Mean number of candidate phases retained by GALAXI's CNNs before passing to DARA as a function of the number of phases in the reference catalog, evaluated on the 61-pattern pristine test set.

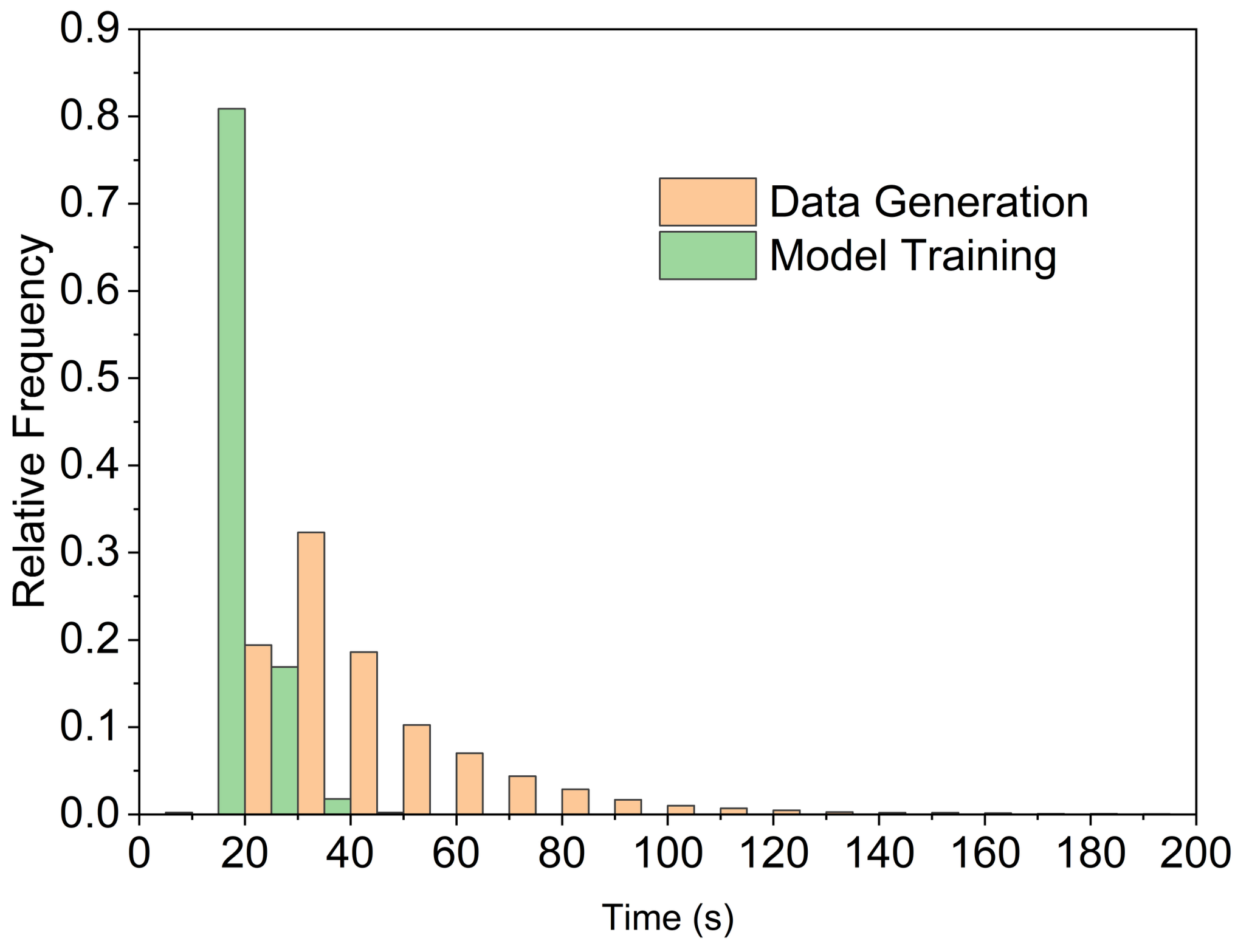

**Supplementary Figure S13.** Distribution of data generation and model training time statistics across the pretrained COD library.

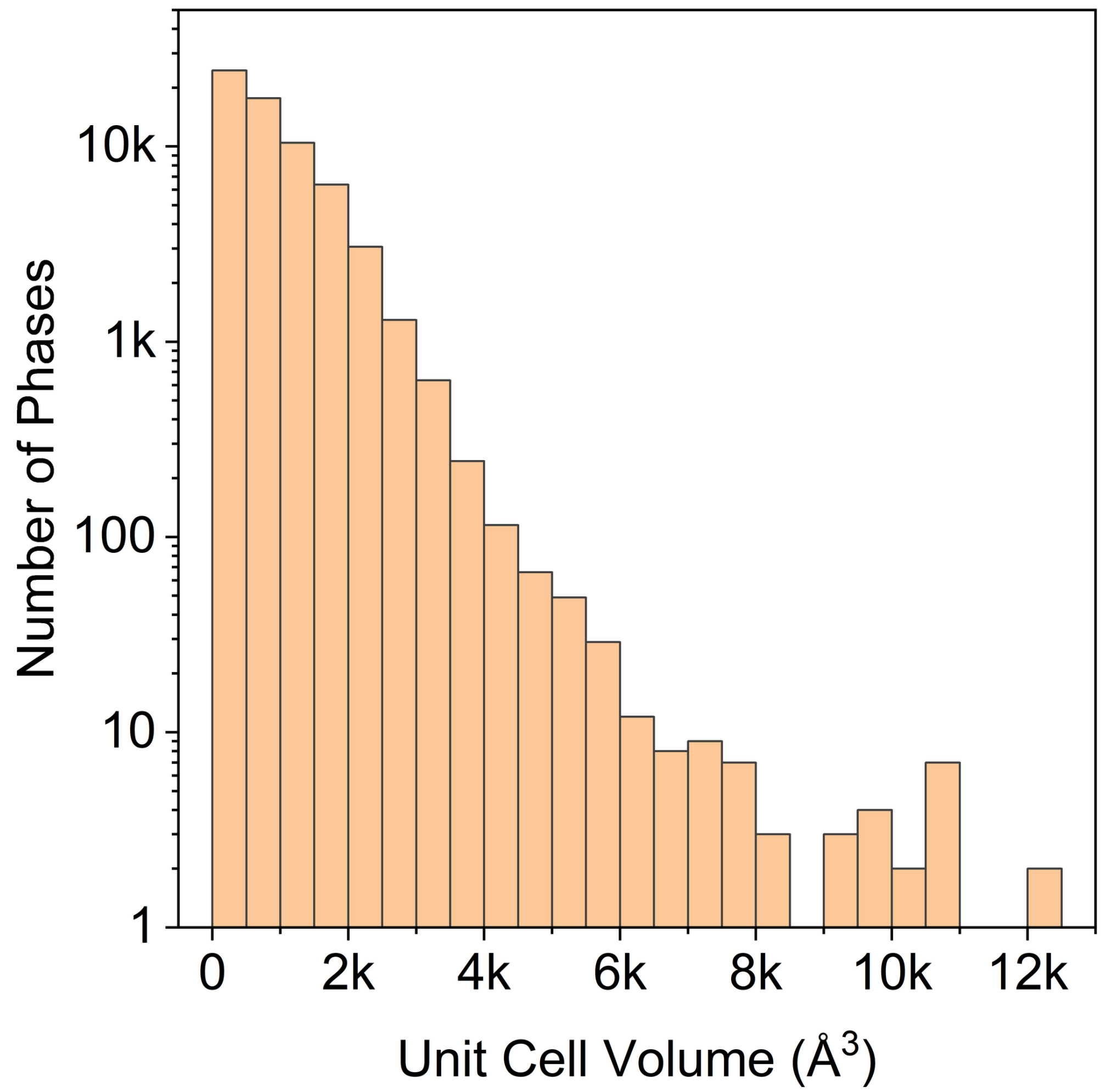

**Supplementary Figure S14.** Distribution of unit cell volume across the pretrained COD library.

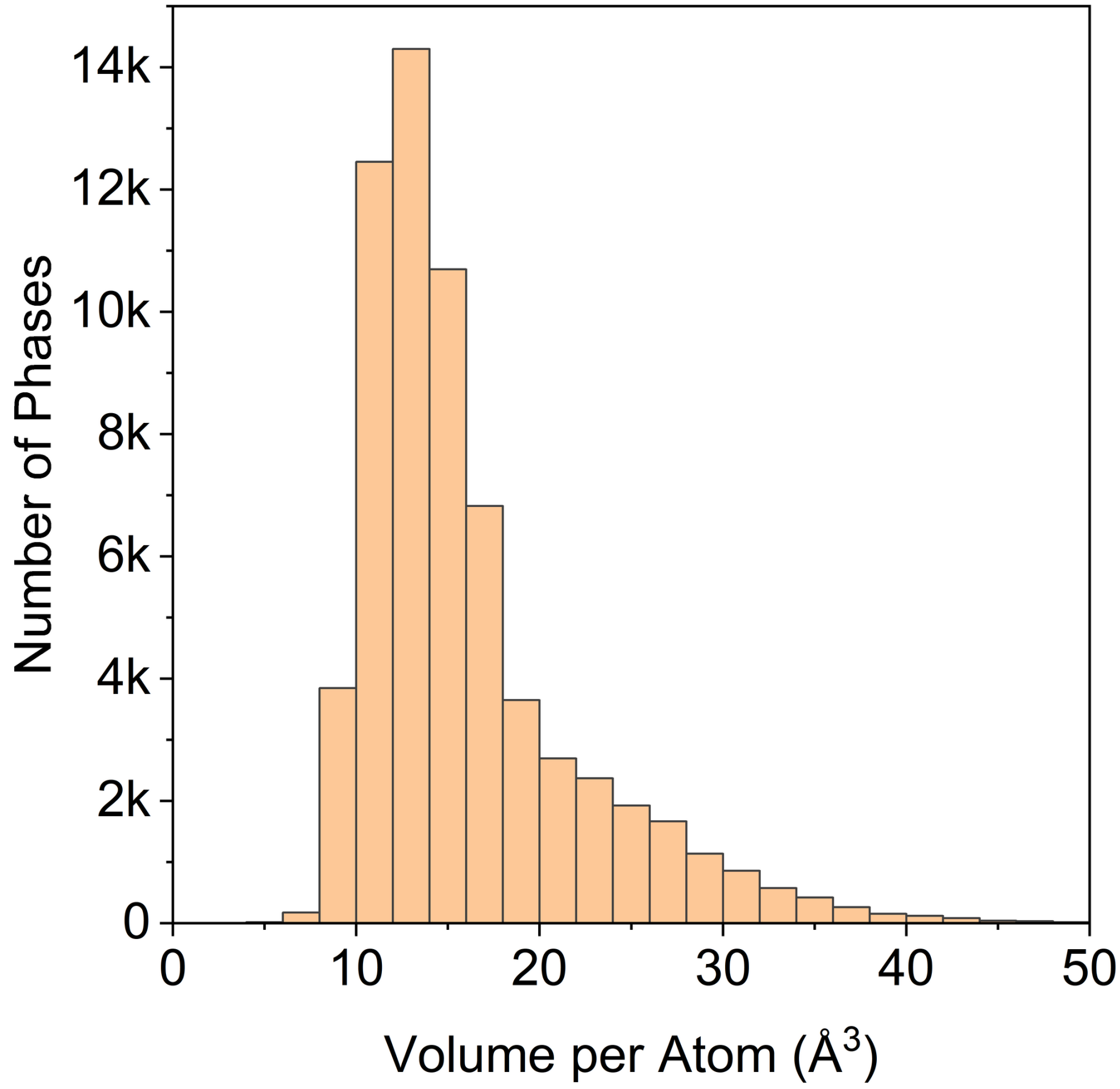

**Supplementary Figure S15.** Distribution of volume per atom across the pretrained COD library.

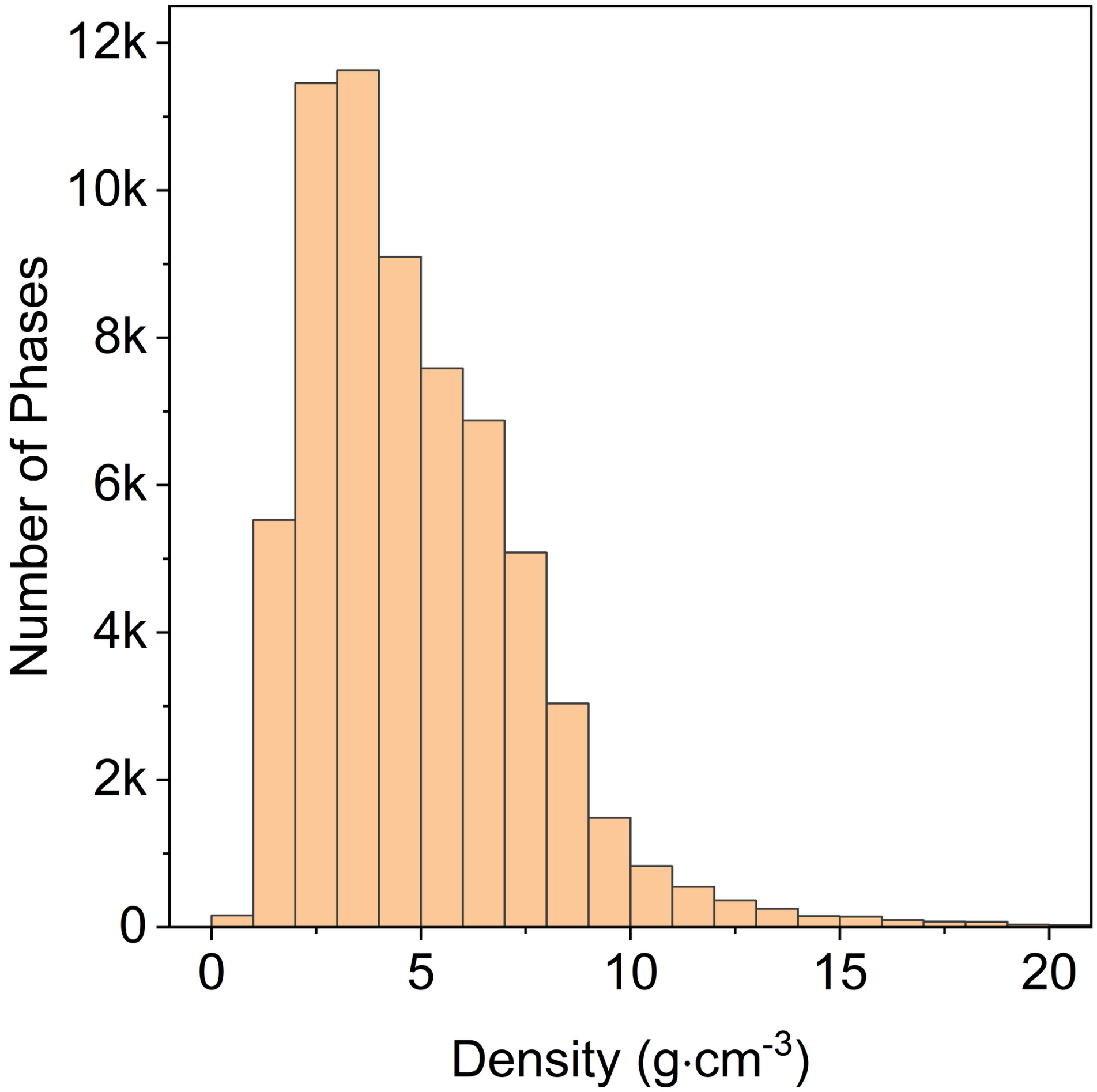

**Supplementary Figure S16.** Distribution of density across the pretrained COD library.

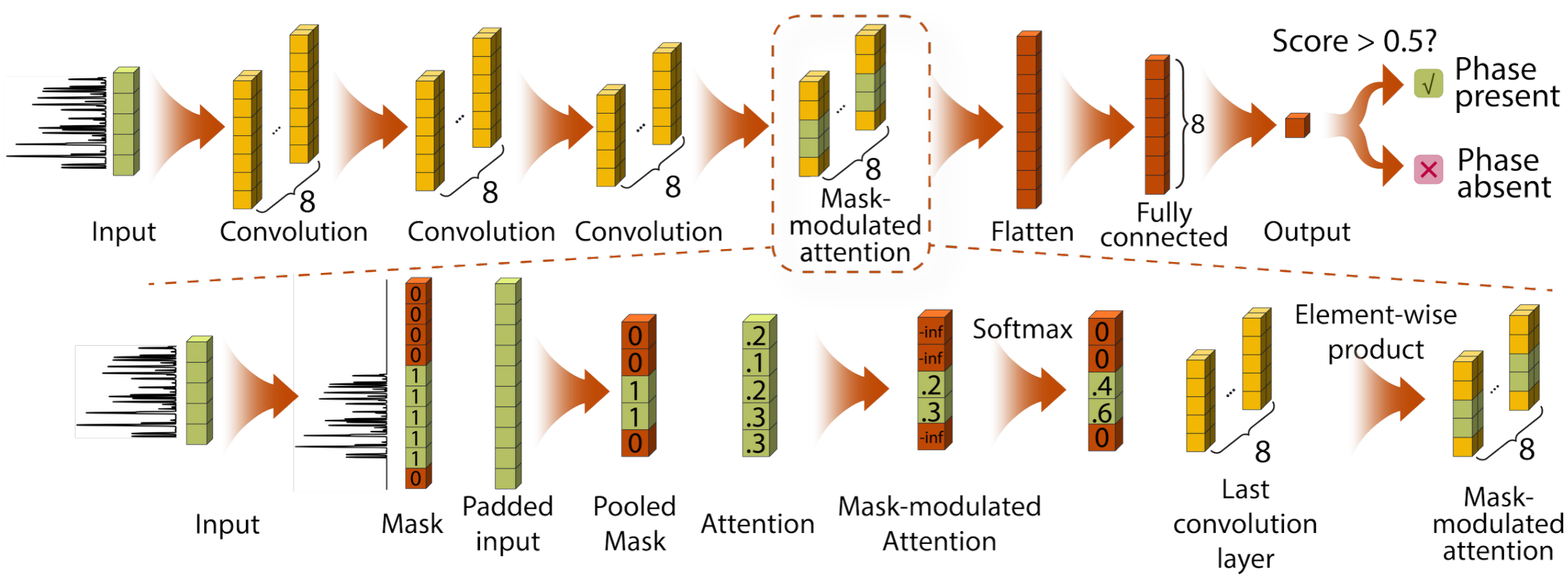

**Supplementary Figure S17.** Schematic of the per-phase CNN architecture, including the mask-modulated attention module.

**Supplementary Table S1. Full one-factor-at-a-time hyperparameter sweep**

Fixed external holdout: 57 experimental patterns and 200 reference candidates, with DARA disabled. Each row represents one completed one-factor-at-a-time trial; bold rows denote the global baseline configuration, repeated in each hyperparameter block for reference. Only parameters showing a discernible trend in micro-F1 are included in the three sub-tables below.

**Table S1.1. Pattern Generation**

| Hyper-parameter | Value/Range | Micro-F1 | Micro Recall | Micro Precision |
|---|---|---|---|---|
| **sample_displacement** | [-0.10, 0.10] | 0.7474 | 0.9730 | 0.6067 |
| | **[-0.50, 0.50]** | **0.7703** | **0.9820** | **0.6337** |
| | [-1.00, 1.00] | 0.7372 | 0.9730 | 0.5934 |
| | [-1.50, 1.50] | 0.7374 | 0.9730 | 0.5902 |
| **uniform_shift** | [-0.05, 0.05] | 0.7308 | 0.8559 | 0.6376 |
| | [-0.10, 0.10] | 0.7333 | 0.8919 | 0.6226 |
| | **[-0.30, 0.30]** | **0.7703** | **0.9820** | **0.6337** |
| | [-0.50, 0.50] | 0.7379 | 0.9640 | 0.5978 |
| | [-1.00, 1.00] | 0.6065 | 0.8468 | 0.4724 |
| **crystallite_size** | [0.1, 450] | 0.7440 | 0.9820 | 0.5989 |
| | [1, 450] | 0.7372 | 0.9730 | 0.5934 |
| | **[5.5, 450]** | **0.7703** | **0.9820** | **0.6337** |

| Hyper-parameter | Value/Range | Micro-F1 | Micro Recall | Micro Precision |
|---|---|---|---|---|
| | [10, 450] | 0.7534 | 0.9910 | 0.6077 |
| | [25, 450] | 0.7883 | 0.9730 | 0.6626 |
| | [50, 450] | 0.8047 | 0.9279 | 0.7103 |
| **lattice_strain** | [0, 0.001] | 0.7754 | 0.9640 | 0.6485 |
| | **[0, 0.005]** | **0.7703** | **0.9820** | **0.6337** |
| | [0, 0.01] | 0.7397 | 0.9730 | 0.5967 |
| | [0, 0.02] | 0.6282 | 0.9820 | 0.4619 |
| | [0, 0.05] | 0.4759 | 0.8468 | 0.3310 |
| **microstrain** | [0, 0.001] | 0.7912 | 0.9730 | 0.6667 |
| | **[0, 0.002]** | **0.7703** | **0.9820** | **0.6337** |
| | [0, 0.003] | 0.7254 | 0.9640 | 0.5815 |
| | [0, 0.004] | 0.7340 | 0.9820 | 0.5860 |
| | [0, 0.005] | 0.7101 | 0.9820 | 0.5561 |
| **target_fraction** | [0.001, 1] | 0.7013 | 0.9730 | 0.5482 |
| | [0.005, 1] | 0.7291 | 0.9820 | 0.5798 |
| | [0.01, 1] | 0.7074 | 0.9910 | 0.5500 |
| | **[0.05, 1]** | **0.7703** | **0.9820** | **0.6337** |

| Hyper-parameter | Value/Range | Micro-F1 | Micro Recall | Micro Precision |
|---|---|---|---|---|
| | [0.10, 1] | 0.7912 | 0.9730 | 0.6667 |
| | [0.20, 1] | 0.8320 | 0.9369 | 0.7482 |
| | [0.35, 1] | 0.8304 | 0.8378 | 0.8230 |
| | [0, 0] | 0.7907 | 0.9189 | 0.6939 |
| | [0, 0.5] | 0.7562 | 0.9640 | 0.6221 |
| | [0, 1.5] | 0.7423 | 0.9730 | 0.6000 |
| **noise_level** | **[0, 3]** | **0.7703** | **0.9820** | **0.6337** |
| | [0, 5] | 0.7543 | 0.9820 | 0.6124 |
| | [0, 10] | 0.7714 | 0.9730 | 0.6391 |
| | [0, 20] | 0.7500 | 0.9189 | 0.6939 |
| | [0.1, 10] | 0.4988 | 0.9009 | 0.3448 |
| | [0.25, 2] | 0.7243 | 0.9820 | 0.5737 |
| **texture** | **[0.5, 1.5]** | **0.7703** | **0.9820** | **0.6337** |
| | [0.75, 1.25] | 0.7458 | 0.9910 | 0.5978 |
| | [1, 1] | 0.7774 | 0.9910 | 0.6395 |

**Table S1.2. Preprocessing**

| | | | | |
|---|---|---|---|---|
| **smoothing_ window_length** | 5 | 0.7365 | 0.9820 | 0.5892 |
| | 11 | 0.7535 | 0.9640 | 0.6185 |
| | **21** | **0.7703** | **0.9820** | **0.6337** |
| | 31 | 0.7010 | 0.9820 | 0.5450 |
| | 41 | 0.6606 | 0.9820 | 0.4977 |
| **snip_iter** | 6 | 0.7692 | 0.9009 | 0.6711 |
| | 12 | 0.7737 | 0.9550 | 0.6503 |
| | **24** | **0.7703** | **0.9820** | **0.6337** |
| | 36 | 0.7914 | 0.9910 | 0.6587 |
| | 60 | 0.7579 | 0.9730 | 0.6207 |
| **noise_ sensitivity** | 1 | 0.7447 | 0.9459 | 0.6140 |
| | 3 | 0.7535 | 0.9640 | 0.6185 |
| | **6** | **0.7703** | **0.9820** | **0.6337** |
| | 9 | 0.7730 | 0.9820 | 0.6374 |
| | 15 | 0.7742 | 0.9730 | 0.6429 |
| **gate_sharpness** | 1 | 0.7910 | 0.9550 | 0.6752 |
| | 3 | 0.7596 | 0.9820 | 0.6193 |
| | **6** | **0.7703** | **0.9820** | **0.6337** |
| | 9 | 0.7397 | 0.9730 | 0.5967 |
| | 15 | 0.6857 | 0.9730 | 0.5294 |
| **Magnification power** | 0.05 | 0.6469 | 0.9820 | 0.4823 |
| | 0.1 | 0.6918 | 0.9910 | 0.5314 |
| | **0.3** | **0.7703** | **0.9820** | **0.6337** |
| | 0.5 | 0.7543 | 0.9820 | 0.6124 |
| | 0.8 | 0.6931 | 0.8649 | 0.5783 |
| | 1.0 | 0.6456 | 0.8288 | 0.5287 |

**Table S1.3. Model Configuration**

| | | | | |
|---|---|---|---|---|
| **learning_rate** | 0.00001 | 0.6688 | 0.9279 | 0.5228 |
| | 0.0001 | 0.7105 | 0.9730 | 0.5596 |
| | 0.0003 | 0.7926 | 0.9640 | 0.6730 |
| | **0.001** | **0.7703** | **0.9820** | **0.6337** |
| | 0.003 | 0.7474 | 0.9730 | 0.6067 |
| | 0.01 | 0.7596 | 0.9820 | 0.6193 |
| | 0.1 | 0.2955 | 0.7387 | 0.1847 |
| **num_points** | 3001 | 0.6119 | 0.9730 | 0.4463 |
| | 5001 | 0.6748 | 0.9910 | 0.5116 |
| | **7001** | **0.7703** | **0.9820** | **0.6337** |
| | 9001 | 0.7224 | 0.9730 | 0.5745 |
| | 11001 | 0.7113 | 0.9099 | 0.5838 |
| **optimizer** | adam | 0.7586 | 0.9910 | 0.6145 |
| | adamw | 0.7660 | 0.9730 | 0.6316 |
| | adagrad | 0.6801 | 0.9099 | 0.5430 |
| | adabelief | 0.7000 | 0.9459 | 0.6158 |
| | **rmsprop** | **0.7703** | **0.9820** | **0.6337** |
| | sgd | 0.7000 | 0.9459 | 0.5556 |
| **pool_size** | [1, 1, 1] | 0.6971 | 0.9640 | 0.5459 |
| | **[2, 2, 1]** | **0.7703** | **0.9820** | **0.6337** |
| | [3, 3, 1] | 0.7754 | 0.9640 | 0.6485 |
| | [4, 4, 1] | 0.7491 | 0.9820 | 0.6056 |
| | [5, 5, 1] | 0.7171 | 0.9820 | 0.5648 |
| **use_mask** | **True** | **0.7703** | **0.9820** | **0.6337** |
| | **False** | 0.6796 | 0.9459 | 0.5303 |